%% file: main.tex
\input{numbers}

\newif\ifsnlayout
\snlayoutfalse            % \snlayouttrue => Springer-Nature submission layout

  \documentclass[fleqn,10pt]{SelfArx}

  \usepackage[english]{babel}
  \usepackage[T1]{fontenc}
  \usepackage{booktabs}
  \usepackage{xcolor}
  \usepackage{orcidlink}

  \usepackage[style=apa,backend=biber,natbib=true]{biblatex}
  \DeclareSourcemap{\maps[datatype=bibtex]{\map{
    \step[fieldsource=url, match=\regexp{\\\_}, replace=\regexp{_}]}}}
  \usepackage{csquotes}
  \usepackage{enumitem}
  \usepackage{balance}
  \definecolor{color1}{RGB}{0,0,90}
  \definecolor{color2}{RGB}{0,20,20}

  \usepackage{hyperref}
  \usepackage{xurl}
  \hypersetup{
    hidelinks,
    colorlinks,
    breaklinks=true,
    urlcolor=color2,
    citecolor=color1,
    linkcolor=color1,
    bookmarksopen=false,
    pdftitle={Measuring Citations to Predatory Journals at Scale: Lessons from arXiv's 2025 BibTeX Files},
    pdfauthor={Andreas Florath}
  }

  \newenvironment{widetable}[1][t]{\begin{table*}[#1]}{\end{table*}}

\newcommand{\citepe}[2]{\citep{#1}}
\newcommand{\citete}[2]{\citet{#1}}

\usepackage[normalem]{ulem}

\newcommand{\PaperTitleText}{Measuring Citations to Predatory Journals at Scale: Lessons from arXiv's 2025 BibTeX Files}
\newcommand{\AffiliationText}{Deutsche Telekom AG, Bonn, Germany. Corresponding author: Andreas Florath, Andreas.Florath@telekom.de}
\newcommand{\KeywordsText}{Predatory Journals, Academic Integrity, Citation Analysis, arXiv, Bibliometrics}
\newcommand{\AbstractText}{Citations to predatory journals have been studied for several years.
  This study measures them in the BibTeX files of arXiv submissions from 2025. We run the
  open-source tool Aletheia-Probe on the \texttt{.bib} files of \CorpusCoveredSubmissions{}
  submissions and compare the cited references with lists of suspected predatory journals.
  Of 8.7 million cited entries, \MultibibCurrentBroadRate\% match at least one such list
  (the broad estimate), and \PrimaryStrictRatePct\% of the 4.3 million
  entries with a confident venue match receive the tool's strict predatory verdict. \SubmissionStrictPct\% of the submissions
  with a cited entry cite at least one reference with this verdict. These figures are operational estimates, not
  validated bounds. Three methodological findings may matter for similar studies. First, the
  lists disagree: the share of a list's matches that yield a predatory verdict when the list itself is ignored ranges from \ReliabilityKscienPublishersLeaveFocalOutPct\% to 0\%. Second, matching venues by name
  alone can attribute evidence to an unrelated venue; an identity audit found this for
  \DerivedAuditMismatchPctOfOriginal\% of the references first labeled strict-predatory. The strict figure above is corrected for this; the broad figure is an upper estimate. Third, raw BibTeX
  counts overstate real reference lists, because some authors bundle whole reference databases.
}

  \JournalInfo{V1}
  \Archive{}
  \PaperTitle{\PaperTitleText}
  \Authors{Andreas Florath\orcidlink{0009-0001-6471-7372}\textsuperscript{1}}
  \affiliation{\textsuperscript{1}\textit{\AffiliationText}}
  \Keywords{\KeywordsText}
  \newcommand{\keywordname}{Keywords}
  \Abstract{\AbstractText}

\begin{document}

\maketitle
\tableofcontents

\section{Introduction}
\label{sec:introduction}

Predatory
journals~\citepe{grudniewicz2019predatory}{p. 210: journals accept articles for fees 'without performing promised quality checks'; p. 212: 'negligent peer review is often a prominent feature of predatory journals'. The comment treats journals and publishers, not conferences. Checked against the PDF on 2026-10-06.}
accept articles for fees without the promised quality checks and often with negligent peer review, and predatory conferences~\citepe{agricola2025fraudulent}{p. 1047, glossary: 'Predatory conference: A conference with weak or no peer review for presentations, poor organization and a focus on making money for the organizers.' Checked against the PDF on 2026-10-06.} have weak or no peer review.
This erodes the reliability of the scholarly
record~\citepe{grudniewicz2019predatory}{p. 211: 'Everyone agrees that predatory publishers sow confusion, promote shoddy scholarship and waste resources.' Checked against the PDF on 2026-10-06.}.
The number of articles in predatory journals grew from an estimated 53{,}000
in 2010 to an estimated 420{,}000 by
2014~\citepe{shen2015predatory}{Abstract (p. 1): predatory journals 'rapidly increased their publication volumes from 53,000 in 2010 to an estimated 420,000 articles in 2014, published by around 8,000 active journals'. The figures are estimates from a stratified sample of 613 journals out of over 11,000 on Beall's list; the article calls its results rough estimates (Limitations). Checked against the PDF on 2026-10-06.}. They have been documented
across several disciplines, from
nursing~\citepe{gabrielsson2020nursing}{Abstract (p. 823): 39 papers by Swedish authors in predatory nursing journals during 2018 and 2019; 'predatory publishing poses a fundamental threat to the development of nursing knowledge'. A case study of one country, not a field-wide estimate. Checked against the PDF on 2026-10-06.} and
radiology~\citepe{aribal2020radiology}{Abstract (p. 498): 66 radiology journals from 27 publishers identified in an updated version of Beall's list (last scan 15 Oct 2019). Counts journals, not articles or citations. Checked against the PDF on 2026-10-06.} to the biomedical
sciences~\citepe{Sharma2018}{Abstract and Introduction (p. 226): narrative review of predatory journals publishing 'worthless biomedical science'; biomedical focus, no data on other disciplines. Checked against the PDF on 2026-10-06.}. A recent report on the mathematical sciences
also names them among new
threats~\citepe{agricola2025fraudulent}{p. 1038 (Sec. 1): paper mills, predatory journals and citation cartels named as new threats; p. 1047 glossary defines predatory journal. The article gives no counts of predatory-journal articles in mathematics, and its glossary says there is no list of predatory journals. Checked against the PDF on 2026-10-06.}. A related phenomenon is the
\emph{hijacked} journal. It copies the identity of a legitimate,
reputable journal to collect submissions under its
name~\citepe{abalkina2021hijacked}{Introduction (p. 2): hijacked journals copy the ISSN and title of a legitimate journal; Argument (p. 7): mimic name, ISSN, reputation and indexing of the original journal and collect fees for fast publication without peer review. Checked against the arXiv v2 preprint on 2026-10-06; volume, pages and DOI of the published version not in the preprint.}. Practical
recommendations~\citepe{agricola2025fighting}{pp. 1179--1183. Sec. 3.1 (p. 1181) lists Retraction Watch with its database, Sec. 3.3 lists Beall's list and Algeria's own list of predatory journals, as 'Internet resources that provide valuable information'. Checked against the PDF on 2026-10-05.} list resources such as
Retraction Watch and the Algerian national list as valuable sources for assessing journals. The Algerian national list is published by Algeria's research directorate~\citep{DGRSDT2024};
it names predatory journals and publishers and defines which journals Algeria recognizes. In this paper, a \emph{predatory list} is a list of journals, conferences or
publishers that its maintainers regard as predatory or have delisted for quality reasons, such as Beall's
List~\citep{bealls_list} or the Algerian national list. A cited reference \emph{matches} a predatory list when its venue appears on that list. A match is a yes or no, not a score.
The \emph{tool} is Aletheia-Probe, open-source software that combines curated
predatory lists, bibliometric databases, retraction records and metadata checks into one transparent,
confidence-scored assessment per venue; a companion paper describes it~\citepe{florath2026aletheia,aletheia_probe}{Abstract; Section 3.3 (pp. 782 to 783): curated backends (DOAJ, Beall's List, PredatoryJournals.org, ministry lists, KSCIEN, Retraction Watch, optional Scopus), Custom List Backend, OpenAlex and Crossref analyzers, Cross-Validator; confidence 0 to 1 as evidence strength (Section 3.5); no source count and no retraction tiers stated. Checked against the submission version on 2026-10-06.}. A \emph{venue} is the journal,
conference or other outlet in which a cited work appeared. A \emph{backend} is one list or database that the tool
queries. The \emph{pipeline} is the whole process of this study:
extracting the references from arXiv, assessing every venue with the tool, and collecting the results.

 This paper adds a large-scale empirical evaluation and stress
test of the tool. It contains an extensive sensitivity analysis and an
add-one evaluation of candidate evidence sources that are not yet part
of the tool (Section~\ref{sec:validation-power}).

By design, we measure what curated predatory lists and bibliometric sources
claim about a venue, and how stable these claims are under
cross-source, leave-one-backend-out and external benchmark checks. We do not
decide whether an individual venue is predatory. That would need a full
validation against independent human judgment, which
Section~\ref{sec:validation-power} discusses. This paper therefore has two aims: to measure how often references
point to venues on predatory lists, and to show how far these numbers can be trusted.

Throughout this paper, a bibliography entry counts as \emph{cited} when the text of the
submission refers to it, not merely when the \texttt{.bib} file lists it. Technically, a \emph{citation} is a
\texttt{\textbackslash cite\{\}}-family command in the submission's
\LaTeX{} source that points to the entry. Some entries appear in the compiled bibliography although our scan
finds no \texttt{\textbackslash cite\{\}} command for them, for example because of
\texttt{\textbackslash nocite\{\}}. Section~\ref{sec:nocite-bbl} describes how we treat them.

The tool processes every raw BibTeX entry in every \texttt{.bib} file
of the \CorpusCoveredSubmissions{} submissions that this study covers: \CorpusAssessedFiles{} files
with \CorpusRawTotalEntries{} raw BibTeX entries. The citation analyses
are restricted to the entries that a submission cites.

We report a \emph{strict} prevalence estimate (the tool's own
confidence-weighted verdict) and a \emph{broad} one (matching at
least one predatory list) side by side, instead of reducing the result to a
single number.

Citation-context analysis is outside our scope. We measure whether a citation points to a predatory or hijacked venue, not whether it
endorses or disapproves of the cited work. 
Prevalence is still worth measuring on its own terms, because even a
citation that does not endorse the cited work may make a venue more visible.

We worked on the entire corpus instead of a curated sample. This
revealed three problems that may matter for other harvested BibTeX
corpora:

\begin{enumerate}[nosep]
  \item \textbf{Cross-source corroboration of the predatory lists varies by an order of magnitude.} We checked each of the nine predatory lists
    against the rest of the tool's evidence
    (Section~\ref{sec:backends}), and the lists behave very differently.
    For the most corroborated list, \ReliabilityKscienPublishersLeaveFocalOutPct\% of its matches still end in a predatory
    verdict when the other backends alone decide. For three small lists, none do. One national
    list~\citep{DGRSDT2024} contributes the most broad-predatory references,
    but without its own vote the rest of the tool does not
    corroborate \DerivedAlgerianMinistryNotRetainedPct\% of them. A single ``percentage of predatory cited references''
    would therefore be misleading without this context.
  \item \textbf{Name-only venue matching can misattribute evidence between unrelated entities.} An identity audit
    (Section~\ref{sec:identity-misattribution}) compared the International
    Standard Serial Number (ISSN) of each ISSN-carrying strict-predatory cited reference with that of the
    venue record behind the predatory vote. It found a mismatch for
    \AuditMismatchRateOfCheckable\% of the references that carry an ISSN, which is
    \DerivedAuditMismatchPctOfOriginal\% of all references first labeled strict-predatory.
  \item \textbf{Bibliography bloat.} Many authors use an entire
    reference database as their paper's bibliography source, most
    visibly the full ACL Anthology~\citep{acl_anthology}, and rely on
    BibTeX to pick the cited subset when the paper is compiled. A na\"ive
    count of raw \texttt{.bib} entries therefore overstates what a paper
    cites by a wide margin. We correct this with a citation log that we
    build from the \LaTeX{} source: a per-file list of the keys that are
    actually cited (Section~\ref{sec:citation-log}).
\end{enumerate}

 Table~\ref{tab:corpus-size} summarizes the corpus; Section~\ref{sec:corpus}
defines its terms.

The rest of the paper reviews related work (Section~\ref{sec:related}), describes the corpus and the
method (Sections~\ref{sec:corpus} and~\ref{sec:methodology}), reports the results
(Section~\ref{sec:results}) and the validation (Section~\ref{sec:validation-power}), and closes with
the discussion, the limitations and the conclusion (Sections~\ref{sec:discussion} to~\ref{sec:conclusion}).

\begin{widetable}[t]
  \centering
  \begin{tabular}{@{}lrr@{}}
    \toprule
    Quantity & Value & \% \\
    \midrule
    arXiv's total 2025 submissions~\citep{arxiv_monthly_stats} & \CorpusTotalSubmissions{} & --- \\
    Distinct submissions covered (this study) & \CorpusCoveredSubmissions{} & \CorpusCoveredSubmissionsPct{} \\
    \quad Submissions with more than one \texttt{.bib} file & \CorpusNMultiBibSubmissions{} & \CorpusMultiBibPct{} \\
    Total \texttt{.bib} files & \CorpusTotalBibFiles{} & --- \\
    \quad Assessed (real bibliography) & \CorpusAssessedFiles{} & \DerivedAssessedPctOfBibFiles{} \\
    \quad Non-bibliography (macro-only) & \CorpusMacroonlyFiles{} & \DerivedMacroonlyPctOfBibFiles{} \\
    \quad Non-bibliography (empty) & \CorpusEmptyFiles{} & \DerivedEmptyPctOfBibFiles{} \\
    Raw BibTeX entries, all assessed files & \CorpusRawTotalEntries{} & --- \\
    \quad in ``mega'' files ($\geq$1{,}000 entries, \DerivedBloatedPctOfPairwiseTotal\% of files) & \CorpusMegaRawEntries{} & \CorpusMegaRawEntriesPct{} \\
    Assessed files, citation log built & \CorpusSidecarFiles{} & \DerivedSidecarPctOfAssessedFiles{} \\
    Assessed files, citation log not built & \CorpusNosidecarFiles{} & \DerivedNosidecarPctDisplay \\
    \midrule
    Raw candidate entries (all entries of files without a log; entries of files with a log only if cited) & \CorpusCandidatePopulation{} & --- \\
    \quad Excluded (no log, but another \texttt{.bib} file's log covers the submission) & \CorpusExcludedEntries{} & \DerivedExcludedPctOfCandidateOnly \\
    \quad Unverified (no log, no other \texttt{.bib} file's log either) & \CorpusUnverifiedEntries{} & \DerivedUnverifiedPctOfCandidate \\
    \quad \textbf{Confirmed-cited (Section~\ref{sec:citation-log}; primary unit of analysis)} & \textbf{\CorpusConfirmedCitedEntries{}} & \DerivedConfirmedCitedPctOfCandidate \\
    \bottomrule
  \end{tabular}
  \caption{Corpus composition. \% by basis: \CorpusCoveredSubmissionsPct{} against arXiv's
    \CorpusTotalSubmissions{} total; \CorpusMultiBibPct{} against the \CorpusCoveredSubmissions{} covered submissions;
    \DerivedAssessedPctOfBibFiles{}, \DerivedMacroonlyPctOfBibFiles{}, \DerivedEmptyPctOfBibFiles{} against the \CorpusTotalBibFiles{} total \texttt{.bib} files;
    \CorpusMegaRawEntriesPct{} against the \CorpusRawTotalEntries{} raw entries; \DerivedBloatedPctOfPairwiseTotal{}, \DerivedSidecarPctOfAssessedFiles{}, \DerivedNosidecarPctDisplay{} against
    the \CorpusAssessedFiles{} assessed files; \DerivedExcludedPctOfCandidateOnly{}, \DerivedUnverifiedPctOfCandidate{}, \DerivedConfirmedCitedPctOfCandidate{} against the
    \CorpusCandidatePopulation{} raw candidate entries.}
  \label{tab:corpus-size}
\end{widetable}

\section{Related Work}
\label{sec:related}

The existence and growth of predatory publishing are well
documented~\citep{grudniewicz2019predatory,shen2015predatory,bohannon2013garbage}.
In biomedical science, authors in predatory journals are mostly young
researchers from developing countries~\citepe{xia2015predatory}{Abstract (p. 2 of the manuscript): 'Those who publish in predatory journals are, for the most part, young and inexperienced researchers from developing countries'; Table 11 gives author origin by region. The study counts publications and citations in biomedical science only. It does not document the growth of predatory publishing. Checked against the author's manuscript PDF on 2026-10-06.}, and the volume of predatory articles differs across
disciplines~\citepe{shen2015predatory}{Fig. 4 (p. 8) and the section 'Distribution over scientific disciplines': 2014 article volumes of predatory journals by discipline in absolute numbers (general about 162,000, engineering 97,000, biomedicine about 70,000). These are volumes, not shares of each field's literature. Checked against the PDF on 2026-10-06.}.
A smaller strand of work asks the question we ask as well: not whether
predatory journals exist, but how often the legitimate literature
actually \emph{cites} them.
The studies below do not share a definition of ``predatory''.
Most identify predatory journals by membership in a list (mostly Beall's
List or Cabell's Predatory Reports) or by the journals of one publisher, and
these lists differ in their criteria~\citepe{strinzel2019blacklists}{Abstract and Results (p. 4): two blacklists (Beall, Cabell) and two whitelists compared; the lists differ in their criteria for quality and the weight given to different dimensions; 72 journals and 42 publishers appear on both a blacklist and a whitelist. Checked against the PDF on 2026-10-06.}.

\citete{frandsen2017citers}{Abstract; Methods: 124 journals from Beall's list of 2014 (random half of the standalone journals, 26 omitted because the name was not unique); citations looked up in Scopus. Results: 1295 citations from 4250 authors to articles published in 2013, window 2013 to 2016. Checked against the accepted manuscript on 2026-10-06.} analyzed authors who cite 124 potentially
predatory journals, using Scopus  to retrieve the citations. \citete{kulczycki2021citation}{Abstract; Methods (p. 5): 65 questionable social-sciences journals selected from Beall's list (37 active, never indexed in WoS) and Cabell's list (37), 74 unique before checks, 65 analysed; 2338 WoS-indexed journals citing them; 3234 cited papers from 25,146 published 2012 to 2018. Checked against the PDF on 2026-10-06.}
profiled the 2{,}338 Web-of-Science-indexed journals whose papers cite 65
questionable social-science journals. \citete{bjork2020frequently}{Abstract and Section 2: 250 articles published in 2014, drawn from journals on Cabell's blacklist (one article per randomly chosen journal), citations counted in Google Scholar over five years. Sample is random over journals, not over articles (Section 2). Checked against the PDF on 2026-10-06.}
tracked five-year Google Scholar citations to 250 articles from
predatory journals. \citete{moussa2020contagion}{Abstract (p. 485): Google Scholar as citation source; 10 predatory marketing journals (11 listed in Table 1, 10 analysed in Table 3); 1246 articles visible in GS received 10,935 citations. 'Citation contagion' is the title term; conclusion: the marketing literature is 'contaminated'. Checked against the PDF on 2026-10-06.} measured Google Scholar
citations to ten predatory marketing journals.
\citete{richtig2023predatory}{Abstract, Methods and Results: Cited-by data for 812 of 5,145 Beall-listed journals in Crossref (15.8 percent), period 2011 to 2018; higher median citation rates for journals also listed in PMC or WoS (Fig 8A). Checked against the full text (Europe PMC) on 2026-10-07.} linked Crossref's Cited-by data to Beall's List. Only 812 of the 5{,}145
Beall-listed journals found in Crossref (15.8\%) provide Cited-by data. Among them, the journals also indexed in PubMed Central or Web of Science show higher citation rates.
\citete{oppenlaender2025citation}{Sections 3.3.6 and 4.2.6: references of CHI articles 1981 to 2024 compared with Beall's List by exact string match after normalisation; 5,126 potential matches reviewed manually; 3,298 citations identified, rising since 2016; one venue, Beall's list only. Checked against the arXiv version of 2025-01-27 on 2026-10-07; published at CHI 2025.} matched the references of ACM CHI papers from 1981 to 2024 against
Beall's List by exact string and reviewed the matches manually. The manual review found
3{,}298 citations, rising since 2016.
\citete{watson2025citation}{Abstract, Methods (pp. 327-329): journals in the 'Medical Sciences' category of one known predatory publisher; 26 listed, 25 included (one excluded, p. 3; the abstract still says 26); 3,671 articles; Web of Science Cited References search. Results (p. 4): 1,151 articles cited by 3,613 citing articles; citing articles rose from 64 in 2014 to 665 in 2022 (10-fold). Checked against the PDF on 2026-10-06.} tracked
Web of Science citations to 3{,}671 articles from the health-sciences
journals of one predatory publisher and found that the number of citing
articles grew roughly tenfold between 2014 and 2022.

\citete{taskin2023worthless}{Abstract; Methods (p. 51): 9,995 citances from 6,706 WoS-indexed papers citing 3,221 questionable articles; 97.3 percent neutral, 1.7 percent positive, 0.8 percent negative (section on citation meaning); tagged by the authors, one person per citance. Checked against the PDF on 2026-10-06.} take a different, content-based approach. Instead of counting references, they classify the sentences around them as positive, neutral or negative towards the cited work.
Of 9{,}995 sentences citing predatory-journal articles, 97.3\% are
neutral. Citing a predatory venue is therefore not by
itself an endorsement, that is, it does not mean that the citing authors approve of the cited work.  \citete{savina2021prevalence}{Abstract and Introduction (arXiv v3): matches Beall's lists of publishers and standalone journals with the Scopus source title list; counts articles in potentially predatory journals by country and subject area; Section 3.1 compares journal-level SNIP and SJR. No analysis of individual citing documents. Checked against the arXiv v3 preprint on 2026-10-06; author name in the published version is Marina (Crossref).} measured
predatory-journal prevalence across the entire Scopus data set at the journal and
country level, and \citete{machacek2022predatory}{Abstract; Introduction: names of potential, possible or probable predatory journals and publishers on Beall's lists mapped to ISSNs (3,293 journals, via Ulrichsweb) and searched in Scopus; 324 journals found with 164,000 articles in 2015 to 2017; share of articles per country for 172 countries in four fields. Counts articles by publishing country, not citing documents. Checked against the PDF on 2026-10-06.} did the same in a
cross-country study of predatory-journal shares among Scopus-indexed
articles.  Both count
articles \emph{published in} predatory venues, not references
\emph{citing} them. This paper looks at the citing side. We examine 8.7 million
bibliography entries that we identified as cited, from \CorpusAssessedFiles{}
assessed BibTeX files.

The studies above show that predatory-journal articles are cited in several
fields. But they differ from this paper
in three ways. First, most start from a curated seed list, often
discipline-specific (tens to low hundreds of journals), and count
citations \emph{to} it. We start from an unselected corpus and ask what fraction
of \emph{its} references resolve to a predatory venue. Second, most
use one citation database to find the citations and one or two predatory lists to decide which journals count, and we are not aware of any that cross-validates
matches against independent evidence sources. Section~\ref{sec:backends} shows that this matters, because
predatory-list corroboration varies widely in our data.
Third, none of the studies discussed here examines arXiv preprints specifically, and none addresses
the bibliography bloat of Section~\ref{sec:bloat}.

Predatory status is better seen as a spectrum than as a yes-or-no
label. \citete{siler2020demarcating}{Abstract: empirical analysis of Cabell's Journal Blacklist shows substantial diversity in types and degrees of predatory publishing; grey journals and publishers occupy borderline niches; predation should be seen as a spectrum, not a binary. Checked against the author's manuscript (docx, 13 Jan 2020) on 2026-10-06; journal, volume, pages from Crossref.} describes varying degrees and types
of illegitimacy among blacklisted publishers. The strict and broad measures in this paper reflect that view in
practice: the predatory lists we use disagree with each other about whether a
venue belongs on a list (Section~\ref{sec:backends}).

A different route uses journal data instead of lists.
\citete{zhuang2025questionable}{Abstract and p. 2: AI method on website design, content and publication metadata; questionable journals defined by the DOAJ best practices; at a balanced threshold the false-positive rate is around 24 percent; the detections are called provisional and subject-area experts should be engaged. Checked against the PDF on 2026-10-12.} train an AI model on website design,
content and publication metadata to identify questionable open-access journals.
At a balanced threshold they report a false-positive rate of around 24\,percent,
and they stress that such detections are provisional and need expert review.

A separate strand of work builds arXiv-scale corpora.
unarXive~\citepe{saier2020unarxive}{Methods (p. 7): arXiv sources flattened with latexpand and parsed with Tralics; reference-section entries stored as database records. Results (p. 11): of 16.9 million matched references, 52.60 percent via Neural ParsCit title, 28.31 percent via DOI (67.1 percent of these heuristically determined, APS patterns), 19.09 percent via arXiv ID; matched references without citation markers are attributed to parsing errors. No mention of nocite. Checked against the PDF on 2026-10-06; issue number from Crossref.} builds a citation graph directly
from arXiv's \LaTeX{} sources. S2ORC~\citepe{lo2020s2orc}{Abstract and Introduction: 81.1M papers from hundreds of publishers and archives; full text for 8.1M open-access papers extracted from PDFs (Science Parse, GROBID; Section 2.1, p. 3); 1.5M additional LaTeX parses from arXiv source where available (Section 2.2, p. 3). Bibliography entries of both parses linked to the corpus (Section 2.4). Checked against the PDF on 2026-10-06; pages and DOI of the proceedings not in the PDF.} works at a larger, multi-publisher scale.
It parses PDFs with tools such as GROBID~\citep{grobid}, and it uses
\LaTeX{} source for the subset of arXiv papers where it is available. Neither describes working from the raw \texttt{.bib} source before BibTeX reduces it to the
cited subset. Section~\ref{sec:bloat} shows how large this difference is
in our corpus.

unarXive links each parsed reference to a record in a paper database, and the figures it reports show how much
of this linking rests on heuristics. Of its matched references,
19.09\% were linked through an arXiv ID and 28.31\% through a Digital Object Identifier (DOI).
Most of these DOIs (67.1\%) were not read from the reference but derived heuristically from
the DOI patterns of American Physical Society journals. The majority (52.60\%) were linked through parsed title strings alone. That is a potential identity risk, but the
figures do not measure misidentification. unarXive also attributes matched references without an in-text citation marker to parsing errors; we found no mention of
\texttt{\textbackslash nocite\{\}} as another cause.

\section{Data and Corpus}
\label{sec:corpus}

\subsection{Corpus construction}
Table~\ref{tab:corpus-size} summarizes the basic counts used throughout
this section.

This study covers only arXiv submissions that provide a \LaTeX{} source
package. arXiv also allows PDF-only submissions, which are entirely
outside our corpus. For each submission we use the arXiv version that
was current when we downloaded the corpus (January to February 2026).
This is not necessarily the version first submitted in 2025.

We extracted the \texttt{.bib} files with a recursive glob over each
submission's \LaTeX{} tar file. This collects every \texttt{.bib} file
that is present and yields \CorpusTotalBibFiles{} files in total. That number
counts files, not submissions: \CorpusNMultiBibSubmissions{} of
\CorpusCoveredSubmissions{} distinct submissions (\CorpusMultiBibPct\%) have more than one
\texttt{.bib} file, and their \texttt{.bib} files make up \CorpusMultiBibFilesPct\% of all
assessed files. Every per-paper statistic in this paper (Section
\ref{sec:results} onward) is therefore computed at the \emph{submission}
level, with all files of a submission pooled, and not at the
\texttt{.bib} file level.

arXiv received \CorpusTotalSubmissions{} submissions in 2025, according to its
published monthly statistics~\citep{arxiv_monthly_stats}. This paper
covers \CorpusCoveredSubmissions{} (\CorpusCoveredSubmissionsPct\%) of them: the submissions with at least one
assessable \texttt{.bib} file. The remaining \CorpusExcludedSubmissions{}
submissions (\CorpusExcludedSubmissionsPct\%) are excluded. They include: PDF-only
submissions without \LaTeX{} source; \LaTeX{} submissions whose
references are written inline with \texttt{\textbackslash bibitem} and
not in a separate \texttt{.bib} file; submissions that use a
non-BibTeX bibliography system; and submissions whose only
\texttt{.bib} files turned out to be macro-only or empty
(Section~\ref{sec:non-bib}). Unless a narrower denominator is stated,
claims in this paper about ``arXiv submissions'' refer to the
\CorpusCoveredSubmissions{} covered submissions, not to arXiv's full 2025 output.
Submission rates that count citations use the \ConcernNPapers{} submissions with
at least one confirmed-cited entry.

The excluded submissions are not a random subset. They differ
systematically in their subject mix. Computer Science makes up
\DisccompinclCsPct\% of the covered submissions but only
\DisccompexclCsPct\% of the excluded ones. Mathematics is the reverse:
\DisccompinclMathPct\% of the covered submissions against
\DisccompexclMathPct\% of the excluded ones. A possible reason is a stronger habit in Mathematics of writing
references inline with \texttt{\textbackslash bibitem}, which we did not
measure. Physics shows the same pattern more
weakly: \DisccompexclPhysicsPct\% of the excluded submissions are
Physics, against \DisccompinclPhysicsPct\% of the covered ones. Seen
from arXiv's side, only about 28\% of Mathematics submissions enter
this corpus, against about 49\% for Physics and 59\% overall. This
matters for the disciplinary comparison in
Section~\ref{sec:disciplinary-breakdown}: the Mathematics group comes
from a small slice (about 28\%) of arXiv's 2025 Mathematics
output.

\subsection{Non-bibliography files}
\label{sec:non-bib}
We checked all \CorpusTotalBibFiles{} files. \CorpusNonbibliographyFiles{} (\DerivedNonbibliographyPctOfBibFiles\%) contain no
citable bibliography entry at all: \CorpusMacroonlyFiles{} contain only
macro definitions (\texttt{@string}, \texttt{@comment},
\texttt{@preamble}), and \CorpusEmptyFiles{} are empty or contain no BibTeX syntax
at all. The remaining \CorpusAssessedFiles{} files (\DerivedAssessedPctOfBibFiles\%) contain real
bibliography entries and were assessed.

\subsection{Bibliography bloat: raw entries overstate real references}
\label{sec:bloat}
The distribution of raw entry counts per file has two clear groups.
\PairwiseNBloatedFiles{} files (\DerivedBloatedPctOfPairwiseTotal\%) have at least 1{,}000 raw BibTeX
entries; we call them \emph{mega files}. Together they hold \CorpusMegaRawEntries{} of the \CorpusRawTotalEntries{} raw
entries (\CorpusMegaRawEntriesPct\%). Manual inspection shows that the dominant recurring families are entire
reference \emph{databases} and not curated bibliographies of single
papers. By raw-entry volume, the full ACL
Anthology~\citep{acl_anthology} dump dominates. Submitting authors use such a
file as the \texttt{\textbackslash bibliography\{\}} source and leave it
to BibTeX to prune it to the cited subset when the paper is compiled. This
is BibTeX's own default behavior, and it matches the definition of a
citation in Section~\ref{sec:introduction}. Section~\ref{sec:citation-log}
looks at the one systematic exception, \texttt{\textbackslash nocite\{\}}.

The largest bibliography in the corpus has \CorpusMaxEntriesPerSubmission{}
entries, but the paper behind it does not have that many genuine references. The paper cites far fewer,
hidden inside a downloaded citation database. We can see this because
we work on the raw \texttt{.bib} source before compilation. Corpora
built from compiled reference lists cannot do this, because BibTeX has already
pruned the database (Section~\ref{sec:related}).

We checked the two dominant recurring families against the tool's
own assessment output: the
ACL Anthology (\MegafamAclPct\% of all mega-file raw entries) and
cryptobib/IACR ePrint (\MegafamCryptobibPct\%). Their \AddoneAclDistinctVenues{} and \BulkCryptobibDistinctVenues{} distinct venues,
respectively, contain no predatory or hijacked classifications and no
adverse retraction-history signals. The only adverse verdicts (\BulkAclFlaggedVenues{} and \BulkCryptobibFlaggedVenues{} venues) are
marked ``suspicious'' (Section~\ref{sec:assessment-logic}). These venues include older or niche historic conferences with
thin modern indexing. None has positive predatory evidence against it.
The remaining \MegafamOtherPct\% of the mega-file entries come from smaller, non-recurring
bibliographies that we did not characterize individually.

\subsection{The citation log}
\label{sec:citation-log}
For every \LaTeX{} submission, arXiv provides its raw source package, but no
ready-made record of which references a paper cites. We build that record
ourselves: the \emph{citation log}. For each submission we scan every
\texttt{.tex} file for \texttt{\textbackslash cite\{\}}-style commands
and match the resulting keys against the submission's own
\texttt{.bib} entries. The log comes only from the raw \texttt{.tex}
and \texttt{.bib} source, never from \LaTeX's compiled \texttt{.aux}
or \texttt{.bbl} output.

\label{sec:log-missing}
Not every \texttt{.bib} file yields a citation log, the per-file list of cited keys. A
\texttt{.bib} file gets a log only if at least one citation key extracted
from the \texttt{.tex} files matches one of its entries. A \texttt{.bib}
file with no match gets no log, whether it is genuinely uncited or the
extraction missed it. We call such a file a \emph{no-citation-log
file}. Of the \CorpusAssessedFiles{} assessed files, \CorpusSidecarFiles{}
(\DerivedSidecarPctOfAssessedFiles\%) yield a log and \CorpusNosidecarFiles{} (\DerivedNosidecarPctDisplay\%) do not. Of the candidate entries,
\DerivedExcludedPctOfCandidate\% (\DerivedExcludedPlusUnverifiedN{} of
\CorpusCandidatePopulation{}) are in these files and have no citation
evidence of their own. Here the candidate population
is the one that Table~\ref{tab:corpus-size} defines: all entries of files without a log, plus the cited
entries of files with a log. We resolve most of
this gap from a structural signal, described next.

\subsubsection{Sibling inference for no-citation-log files}
\label{sec:sibling-inference}
Some submissions bundle more than one \texttt{.bib} file. For these, we
pool every citation key that is used anywhere in the submission's
\LaTeX{} source. Then we check each of the submission's files (its
\emph{siblings}) against that pool. A file without a log of its own
might simply be unused, such as a bundled reference database. Or the
extraction might have failed to parse it. If a sibling has a log and none of the file's
keys is among the keys cited in the submission, we assume that the file is unused and call its entries \emph{excluded}, because nothing in the submission cites them. The inference covers \CorpusNosidecarSiblingFilesN{} of
the \CorpusNosidecarFiles{} no-citation-log files (\CorpusNosidecarSiblingPct\%). As a check, the combined signal of Section~\ref{sec:nocite-bbl} confirms cited entries in only \SisfSibNPromotedFiles{} of these files: \SisfSibNPromotedEntries{} entries in \SisfSibNPromotedSubmissions{} submissions.
None of them is strict-predatory, and one is broad-predatory. For the remaining
\CorpusNosidecarIsolatedFilesN{} (\CorpusNosidecarIsolatedPct\%), the isolated files, no sibling yields a log either. We do not
guess for these. We report them separately as their own category
(Section~\ref{sec:no-sidecar-sensitivity}).

This gives three groups, which together make up the whole candidate
population (Table~\ref{tab:corpus-size}). Entries
confirmed as cited make up the \CorpusConfirmedCitedEntries{} confirmed-cited entries:
\ConfpopNOwnLog{} are confirmed by their file's own citation log,
\ConfpopNMissedKeys{} are keys that the log missed but that the combined signal of Section~\ref{sec:nocite-bbl} recovers,
and \ConfpopNPromotedNoLog{} are added in files without a log (\ConfpopNPromotedNoLogIsolated{} in isolated files, \ConfpopNPromotedNoLogSibling{} in sibling-inferred ones)
(Sections~\ref{sec:citation-log-accuracy} and~\ref{sec:nocite-bbl}). Together, the files inferred to be unused via a sibling
hold \CorpusExcludedEntries{} excluded entries, \DerivedExcludedPctOfCandidateOnly\% of the candidate population. The isolated files, with neither a log of their own nor a sibling's,
hold \CorpusUnverifiedEntries{} unverified entries. All statistics in
this paper use the \CorpusConfirmedCitedEntries{} confirmed-cited entries as the primary unit of
analysis, not the \CorpusRawTotalEntries{} raw BibTeX entries and not the larger candidate
population, unless we say otherwise.

\subsubsection{Accuracy of the citation log}
\label{sec:citation-log-accuracy}
As a check, we compared the citation log with arXiv's own compiled
bibliography (\texttt{.bbl}) for the \CorpusSidecarFiles{} files for which we built
a log. The two agree closely. \BblaccuracyRecallPctShort{}\% of the keys that
\texttt{.bbl} confirms are also in the citation log, and
\BblaccuracyPrecisionPctShort{}\% of the keys in the citation log are also
confirmed by \texttt{.bbl}. Part of the small remaining gap comes from
two citation-macro variants that the log does not cover:
\texttt{\textbackslash combocite} and apacite's
\texttt{\textbackslash shortcite}.

A gap of this kind is expected. \LaTeX{} is a complete
macro-programming language, so a paper can define its own citation
commands, and catching every such case would mean reimplementing
\LaTeX{} and its packages. That is not a realistic target. We do not try to
support every citation-macro variant directly. The final analysis
instead recovers keys that the citation log misses but the combined signal finds (Section~\ref{sec:nocite-bbl}). Of the \MissedkeysNPopulation{} entries recovered in files that have a log, \MissedkeysCurrentStrictAfterN{} are strict-predatory and \MissedkeysCurrentBroadAfterN{} broad-predatory. We keep the citation log as the primary \emph{per-file}
record because it applies uniformly, even where no compiled
\texttt{.bbl} is available.

\subsubsection{\texttt{\textbackslash nocite\{\}} in the compiled bibliography}
\label{sec:nocite-bbl}
As anticipated in Section~\ref{sec:introduction},
\texttt{\textbackslash nocite\{\}} can sweep a \texttt{.bib} entry into
the compiled bibliography without a genuine in-text citation.
\NociteBblOnlyTotal{} keys appear in the compiled bibliography but not in our scan of the \LaTeX{}
source for cite commands. Of these, \NociteBblOnlyAttributable{}
(\NociteBblOnlyAttributablePct{}\%; not every key has a \texttt{.bib} entry) trace to
\texttt{\textbackslash nocite\{\}}. We treat the remaining
\NociteBblOnlyNonAttributable{} keys as genuine citations that the citation log
missed. The combined signal combines the compiled bibliography's keys
with those of a second, more complete scan of the \LaTeX{} source for cite commands, and it cannot tell
the two cases apart, so the \texttt{\textbackslash nocite\{\}}-attributable entries stay in the primary population. It promotes \NociteNPromotedTotal{} entries to
confirmed-cited status: \ConfpopNMissedKeys{} in files that have a citation
log and \ConfpopNPromotedNoLog{} in files that have none. Of these
\NociteNPromotedTotal{} entries, \NociteNPromotedAttributable{}
(\NocitePctOfPromotedAttributable{}\%) are
\texttt{\textbackslash nocite\{\}}-attributable.
Section~\ref{sec:nocite-sensitivity} shows that excluding them from the
confirmed-cited population does not change the headline rate.

\section{Methodology}
\label{sec:methodology}

\subsection{Pipeline and infrastructure}
Assessment was performed with Aletheia-Probe~\citep{aletheia_probe}. It
queries 22 evidence backends for every venue. The backends run
against locally hosted data and not against live remote APIs, because
many APIs allow only a limited number of requests per time.

The 22 backends play four roles (Table~\ref{tab:backend-roles}). Nine are predatory lists that vote \emph{predatory}. Eight are legitimate lists that vote \emph{legitimate} when they find the venue. Three are analyzers: they match no list but analyze a venue's publication and citation patterns and the quality of its metadata, for example publication growth, citation ratios and the share of articles with ORCID, funding or license information, and they can point either way. Two backends do not vote: KSCIEN's hijacked-journal list and Retraction Watch (Section~\ref{sec:assessment-logic}). UGC-CARE (University Grants Commission, Consortium for Academic and Research Ethics, India) publishes its journals in two groups, Group I and Group II. Its cloned-journal lists and the list of journals delisted from Group II count as predatory lists, and the included journals named on the two clone pages count as legitimate lists.

\begin{table}[htp]
  \centering
  \footnotesize
  \begin{tabular}{@{}p{0.2\linewidth}p{0.55\linewidth}p{0.14\linewidth}@{}}
    \toprule
    Role & Backends & Votes \\
    \midrule
    Predatory lists (9) & Beall's List~\citep{bealls_list}; \citet{predatoryjournals}; the Algerian national list~\citep{DGRSDT2024}; KSCIEN~\citep{kscien} (standalone journals, conferences, publishers); UGC-CARE~\citep{ugc_care} (cloned Group I, cloned Group II, delisted Group II) & predatory \\
    Legitimate lists (8) & DOAJ~\citep{doaj}; Scopus~\citep{scopus}; DBLP~\citep{dblp}; PubMed~\citep{pubmed}; \citet{core} (journals and conferences, two backends); UGC-CARE included journals (Group I, Group II) & legitimate \\
    Analyzers (3) & \citet{openalex}; \citet{crossref}; \citet{opencitations} & either \\
    No vote (2) & KSCIEN hijacked journals; \citet{retraction_watch} & none \\
    \bottomrule
  \end{tabular}
  \caption{The 22 backends by role. The number in brackets is the number of backends in the role.}
  \label{tab:backend-roles}
\end{table}

All 22 backends use data sources that are free of charge. Where the underlying
product is commercial, we use only a freely available subset. For
Scopus this is the Source List with title, ISSN and CiteScore. Only titles with the status active are loaded, so titles that Scopus discontinued for publication concerns are not among them. No paid
subscription is needed. For the same reason we exclude Cabells'
Predatory Reports~\citep{cabells_predatory_reports}, a commercial
predatory-journal database with over 20{,}000 listed journals.

Each backend is queried separately. This does not mean that their
outputs are statistically independent. Section~\ref{sec:backends}
discusses the correlation that we cannot rule out between sources that
may share underlying indexing coverage.

The evaluation reported here ran the tool over every one of the
\CorpusRawTotalEntries{} raw \texttt{.bib} entries, including entries not
identified as cited. Entries without an extractable venue name or with an
unsupported entry type receive no venue assessment. It took
approximately 9.5 days (2026-03-31 to 2026-04-09). The per-citation
dataset of this paper uses these results, restricted to
confirmed-cited entries (Section~\ref{sec:citation-log}). The
evaluation ran on a development state of Aletheia-Probe (commit 9b5869b,
17 March 2026), with small local
adjustments that do not change the scoring. All of them are included in
release 0.10.0.

\subsection{Assessment logic}
\label{sec:assessment-logic}
For each entry, the backends query with whatever identifying
information is available: the ISSN when present and supported, otherwise the venue
name. The nine predatory lists all match by venue name, and four of
them also by ISSN (Appendix~\ref{sec:appendix-identity-forensics}). If no backend returns a confident match, the tool retries
with a name that it resolves from an acronym/alias table. This is how
short forms are matched against full venue names; some variants stay
unresolved (Appendix~\ref{sec:appendix-sensitivity}).

Each backend reports \texttt{found} or \texttt{not\_found}. When it
finds the venue, it also reports an assessment (\texttt{predatory},
\texttt{legitimate}, or one of several other status values) and its own
per-query \emph{confidence} score between 0 and 1. Each backend
computes this confidence in its own way.

Two of the 22 backends do not count toward whether a venue is
predatory at all: KSCIEN's hijacked-journal list and Retraction Watch.
They feed neither the confidence-weighted predatory/legitimate vote nor
the broader ``matching at least one predatory list'' count
(Section~\ref{sec:strict-broad}). The reason is that both answer a
different question than ``is this venue predatory''. The
hijacked-journal list marks venues whose identity has been hijacked,
not venues that are predatory. Retraction Watch records whether a venue
has ever published a retracted article, which is a quality signal and
not a predatory-journal one. We still report both separately in
Section~\ref{sec:results}, just not folded into either count.

Each backend can be given an individual weight that sets how much
influence it has on the overall result. In this paper all vote-eligible
backends have the same configured weight. Their realized influence still
differs, through their roles, confidence scores, coverage and conflicts.

A \texttt{predatory} verdict needs a match with at least one of the nine predatory lists, and that evidence must win a
confidence-weighted comparison against any legitimate evidence. In this comparison, every backend that votes adds
its confidence times its weight to its own side, and the predatory side wins only if its sum is larger. There is no
further threshold.
Predatory evidence from the analyzers alone, without such a match, leads to \texttt{suspicious} and never to
\texttt{predatory}. The voting and conflict-resolution logic is
described in the software repository~\citep{aletheia_probe}. The
companion paper describes the principles~\citepe{florath2026aletheia}{Section 3.5: sources reported individually, conflicts reported, confidence lowered by conflicts, sparse and heuristic-only evidence; no voting rules described. Checked against the submission version on 2026-10-06.}.

Retrieving each backend's raw evidence (ISSN and DOI resolution, venue
matches, per-query confidence) is designed to be deterministic with
the same input data. Combining this evidence into one verdict with the
rule above is an interpretive decision under a documented rule. It is
not an independently established fact about a venue's editorial
practices.

\subsection{Two prevalence estimates: strict and broad}
\label{sec:strict-broad}
For the headline prevalence estimates we report two forms. The strict estimate
uses the tool's confidence-weighted verdict, its top-level
assessment of the venue as \texttt{predatory}, computed with the identity correction of
Section~\ref{sec:identity-misattribution}. We call it the strict verdict, although a strict label does not always mean that
multiple sources agree (see Sections~\ref{sec:backends},
\ref{sec:leave-one-out} and~\ref{sec:strict-alternatives}). The broad
estimate counts any cited reference that matches at least one predatory list, whether or not the remaining evidence agrees. Broad is the more inclusive
measure, strict the more selective one. Neither is a validated upper
or lower bound on the true prevalence, and each alone would give an
incomplete picture of how many cited references touch \emph{some}
amount of documented concern. Section~\ref{sec:backends} quantifies how
far apart the two numbers are and why.

\subsection{Denominators for strict and broad}
\label{sec:denominators}
Strict and broad are not the same kind of quantity. Using one
denominator convention for both would be a source of error, so we state
the conventions explicitly.

The confidence-weighted vote reaches a decision only once at least one
backend returns a sufficiently confident match. When it does, the
result is one of four \emph{resolved} outcomes: \texttt{legitimate},
\texttt{predatory}, \texttt{suspicious} or \texttt{insufficient\_data}.
An entry with no confident match gets no such outcome and is
\texttt{unknown}. Broad status is different. It is a raw count of
matches with predatory lists and does not depend on whether the vote resolves. An
entry can therefore be broad-predatory while its strict assessment is
still \texttt{unknown}.

Table~\ref{tab:denominators} lists every population this implies. The
first three entry populations are nested, each inside the one above, and
the last row gives the number of unique submissions. Confirmed-cited is this paper's full
primary population (Table~\ref{tab:corpus-size}). Eligible keeps only
the entries for which a predatory verdict makes sense at all: those
with an extractable venue name and a supported BibTeX entry type
(\DerivedAssessedPctDisplay\% of the confirmed-cited entries). It leaves out preprints
(\DerivedPreprintPctDisplay\%), unsupported entry types such as \texttt{@book}
(\DerivedUnsupportedTypePctDisplay\%), and entries without a recoverable venue name (\DerivedNoVenuePctDisplay\%).
Resolved-eligible goes one step further. It
keeps the eligible entries for which some backend returned a confident
match, so the vote could reach a real verdict instead of
\texttt{unknown}. Section~\ref{sec:headline-prevalence} reports strict
and broad prevalence under each of these denominators.

\begin{table}[htp]
  \centering
  \footnotesize
  \begin{tabular}{@{}lr@{}}
    \toprule
    Population & $n$ \\
    \midrule
    \textbf{Confirmed-cited (this paper's unit of analysis, Table~\ref{tab:corpus-size})} & \textbf{\CorpusConfirmedCitedEntries{}} \\
    \quad Eligible (assessable venue \& entry type) & \EligiblePopulation{} \\
    \quad\quad Resolved-eligible (a confident match was found) & \ResolvedEligiblePopulationNew{} \\
    Unique submissions among confirmed-cited entries & \DiscsubcountNTotal{} \\
    \bottomrule
  \end{tabular}
  \caption{Population sizes used as prevalence denominators: cited-reference
    entry populations and unique submissions.}
  \label{tab:denominators}
\end{table}

We take resolved-eligible as the primary strict denominator. It
answers the natural question: of the cited references that we could
classify, what fraction is predatory? This avoids diluting the
rate with a large unresolved group that, by construction,
contributes nothing to its numerator. The predatory share among \texttt{unknown}
entries is not known, and Section~\ref{sec:unknown-sensitivity} shows how the
count changes under assumed rates.

Broad has no resolved-only variant. A match with a predatory list does not need the
vote to resolve, so restricting the broad denominator to
resolved-eligible entries would drop broad-predatory but unresolved cited
references from the numerator for no good reason. Broad is therefore
reported over all confirmed-cited entries throughout.

\begin{widetable}[t]
  \centering
  \begin{tabular}{@{}lrr@{}}
    \toprule
    Assessment & Count & \% \\
    \midrule
    Legitimate & \HeadlineLegitimateN{} & \HeadlineLegitimatePct{} \\
    Not applicable (preprint/no-venue/unsupported) & \HeadlineNotApplicableN{} & \HeadlineNotApplicablePct{} \\
    Unknown (no confident match) & \HeadlineUnknownN{} & \HeadlineUnknownPct{} \\
    Suspicious & \HeadlineSuspiciousN{} & \HeadlineSuspiciousPct{} \\
    \textbf{Predatory (strict)} & \textbf{\StrictCount{}}$^a$ & \textbf{\MultibibCurrentStrictRate{}} \\
    Insufficient data & \HeadlineInsufficientDataN{} & \HeadlineInsufficientDataPct{} \\
    \midrule
    Total & \HeadlineTotal{} & \DerivedHeadlineRowsSumPct{} \\
    \midrule
    \textbf{Predatory (broad: $\geq$1 predatory list)} & \textbf{\BroadCount{}}$^a$ & \textbf{\MultibibCurrentBroadRate{}} \\
    Hijacked (excluded from the above by design) & \HeadlineHijackedN{} & \HeadlineHijackedPct{} \\
    \bottomrule
  \end{tabular}
  \caption{Final-assessment distribution, all \HeadlineTotal{} confirmed-cited
    entries. $^a$Predatory (strict/broad) are operational counts, not validated counts or bounds. The broad count is corrected for identity mismatches only through the strict references (Section~\ref{sec:identity-misattribution}). The rows above the Total row sum to the total. The strict percentage here uses all confirmed-cited entries as denominator; the headline rate is in Table~\ref{tab:rates}.}
  \label{tab:headline}
\end{widetable}

\subsection{Reproducibility}
\label{sec:reproducibility}
We did not repeat the whole measurement. We re-evaluated a small sample
of already-assessed files about 3.5 months after the main run. The final
assessments differed in less than 1\% of the entries. The sample is not
random and leaves out the largest and smallest files, and it holds only
\ReevalsampleStrictMainN{} entries with a strict \texttt{predatory} verdict,
none of which changed. It therefore says little about the stability of the
strict count. A likely reason for the differences is that the tool's local
database is extended and changed while a run is in progress, so the
second run did not start from the same state as the first. We treat this
as the relabeling rate seen in this sample, not as a corpus-wide noise
floor.

\subsection{Use of AI tools}
\label{sec:ai-tools}
Several AI systems were used to prepare this paper. Claude (Anthropic)
helped to draft and revise the text, to restructure sections, to fix
\LaTeX{} layout problems, to check bibliographic references against
independent sources, and to write and run parts of the analysis scripts.
ChatGPT (OpenAI) and Gemini (Google) served as additional reviewers of
drafts, and local open-weight models (for example Qwen3-Coder) as
coding assistants. The author reviewed all AI-assisted text and output,
checked the references and the reported numbers against the underlying
data, and is fully responsible for the content. No AI tool is listed as
an author.

\section{Results}
\label{sec:results}

\subsection{Headline prevalence}
\label{sec:headline-prevalence}
This subsection gives the main result: how many cited references point to predatory venues, and how the answer changes with the denominator.

Table~\ref{tab:headline} reports the full final-assessment distribution
over the \CorpusConfirmedCitedEntries{} confirmed-cited entries. The rows above the Total row divide that population and sum to 100\%. The two rows below are
non-exclusive criteria that overlap with those rows. \texttt{Legitimate} is the
single largest outcome (\HeadlineLegitimatePct{}\%), with \texttt{unknown} at \HeadlineUnknownPct{}\%.

Table~\ref{tab:rates} reports strict and broad prevalence under each
of the denominator conventions defined in
Section~\ref{sec:denominators}. In that table, each estimate's
numerator is fixed --- \StrictCount{} confirmed-cited entries whose
confidence-weighted verdict resolved to \texttt{predatory} for every
row labelled strict, and \BroadCount{} confirmed-cited entries matching at least one
predatory list for every row labelled broad --- only the denominator (the population we divide by) changes
between rows. Primary marks the denominator we treat as each
estimate's headline figure; secondary rows use the identical
numerator under an alternative, equally valid denominator convention,
reported for transparency rather than adopted as the headline number.

\begin{table}[htp]
  \centering
  \footnotesize
  \begin{tabular}{@{}lrrr@{}}
    \toprule
    Estimate & Num. & Denom. & \% \\
    \midrule
    \textbf{Strict, resolved-eligible (primary)} & \StrictCount{} & \ResolvedEligiblePopulationNew{} & \textbf{\PrimaryStrictRatePct{}} \\
    Strict, eligible (secondary) & \StrictCount{} & \EligiblePopulation{} & \SecondaryEligibleRatePct{} \\
    Strict, all confirmed-cited (secondary) & \StrictCount{} & \CorpusConfirmedCitedEntries{} & \MultibibCurrentStrictRate{} \\
    \textbf{Broad, all confirmed-cited (primary)} & \BroadCount{} & \CorpusConfirmedCitedEntries{} & \textbf{\MultibibCurrentBroadRate{}} \\
    \bottomrule
  \end{tabular}
  \caption{Strict and broad predatory-citation prevalence across the
    cited-reference denominator conventions in Table~\ref{tab:denominators}.}
  \label{tab:rates}
\end{table}

We also count whole papers. Here the unit is the arXiv
\emph{submission}, not the \texttt{.bib} file, because one submission
can bundle several (Section~\ref{sec:corpus}). Of the
\DiscsubcountNTotal{} submissions with at least one confirmed-cited
entry, \DiscsubcountNStrict{} (\SubmissionStrictPct\%) cite at least
one strict-predatory venue, and \DiscsubcountNBroad{}
(\SubmissionBroadPct\%) cite at least one broad-predatory venue.

A third, wider criterion combines all three of our signals. A
submission shows ``any documented concern'' if it cites a
broad-predatory venue, a venue with a hijacked identity, or a venue
in the moderate, high or critical retraction-risk tier
(Section~\ref{sec:retraction}). \DerivedConcernPctOfPapers\% of
submissions (\ConcernNAnyFlag{} of \ConcernNPapers{}) meet it.

Retraction risk drives most of that number. Of the submissions that meet it, retraction risk is involved in
\ConcernPctRetractionOfFlagged\% and is the only reason for
\ConcernRetractionOnlyPct\%; broad-predatory citations are involved in
\ConcernBroadTotalPct\% and the only reason for
\ConcernBroadOnlyPct\%; hijacked identities, \ConcernHijackedTotalPct\%
and \ConcernHijackedOnlyPct\%. Read alone,
\DerivedConcernPctOfPapers\% sounds like a predatory-citation
statistic; it is mostly a retraction-risk one.

The denominator here is \ConcernNPapers{} submissions, not all
\CorpusCoveredSubmissions{} we assessed: \DerivedSubmissionsWithoutConfirmedCitedN{} submissions cite nothing
we could confirm, so they hold no reference we could match either way.
Most of the \DiscsubcountNBroad{} broad-predatory submissions cite
exactly one broad-predatory venue (\SubbroadhistNOne{} of them), and
\SubbroadhistNFourplus{} cite four or more.

\subsection{The predatory lists disagree: cross-source corroboration}
\label{sec:backends}
The predatory lists disagree with each other. How far the rest of the tool backs a list's matches
ranges from \ReliabilityKscienPublishersLeaveFocalOutPct\% for the best-corroborated list to 0\% for three small
lists, and for the dominant national list it is \ReliabilityAlgerianMinistryLeaveFocalOutPct\%. Two measures, defined below,
show this.

Of the \BroadCount{} broad-predatory
cited references, \BackendsSplitPctExactlyOne\%\ match exactly one predatory list and the rest two or more. 
The first measure is a leave-focal-out check (Table~\ref{tab:reliability}). It differs from the
leave-one-backend-out analysis of Section~\ref{sec:leave-one-out}, which asks how much dropping a backend
moves the headline count. Here, for each reference that matches a list, we drop that list's own vote, rerun the
confidence-weighted decision rule over every other backend's stored
result, and report how often the verdict is \texttt{predatory}.

This is a \textbf{cross-source corroboration rate}: how often the other
sources agree with a list. It does not say how often the list is right,
because the rest of the tool is no ground truth: those are other bibliometric and
indexing sources, which may correlate with one another, may measure
indexing coverage rather than editorial quality, and may serve purposes
unrelated to predatory-journal detection
(Section~\ref{sec:limitations}).

Table~\ref{tab:reliability} reports a second measure beside it:
\emph{Second vote}, the share of a list's matches that carry a
predatory-side vote from somewhere else too. That other vote can come
from a second predatory list or from another kind of backend. A low
second-vote share means a list matches alone: nothing else in the
tool points the same way. A high one means other sources see
something too, but not necessarily enough, since the
confidence-weighted decision rule can still refuse a
\texttt{predatory} verdict on that evidence by itself. The Algerian
national list matches most of its entries alone, at a second-vote share of
\ReliabilityAlgerianMinistryCoFlagPct\% which is the lowest in
Table~\ref{tab:reliability}.

The two measures can pull far apart, and the gap is the interesting
part. The KSCIEN publishers list and predatoryjournals.org score high on
both, at \ReliabilityKscienPublishersLeaveFocalOutPct\% and
\ReliabilityPredatoryjournalsLeaveFocalOutPct\% leave-focal-out: drop
either list's own vote and its matches mostly stand regardless. Three lists
sit at the opposite end, all at \ReliabilityKscienPredatoryConferencesLeaveFocalOutPct\%:
the KSCIEN predatory-conferences list and the two UGC-CARE lists in
Table~\ref{tab:reliability}'s last two rows. Not one of their matches
survives losing its own vote. Their second-vote shares
are nonetheless \ReliabilityKscienPredatoryConferencesCoFlagPct\%,
\ReliabilityUgcCareDelistedGroupCoFlagPct\% and
\ReliabilityUgcCareClonedCoFlagPct\%.

For most of these entries another source also votes on the predatory
side, but the replayed rule does not return predatory on that evidence
alone.  These three lists are small
(\ReliabilityKscienPredatoryConferencesFlags{}, \ReliabilityUgcCareDelistedGroupFlags{}
and \ReliabilityUgcCareClonedFlags{} matches). We did not split their second-vote support by
source type.

The Algerian national list~\citep{DGRSDT2024} is a related case at a
much larger scale. It matches \AlgerianBroadSharePct\% of all broad-predatory cited references (\ReliabilityAlgerianMinistryFlags{} of \BroadCount{}), and the
rest of the tool does not corroborate
\DerivedAlgerianMinistryNotRetainedPct\% of those matches once the list's own vote is removed.

One venue accounts for about half of that. \JscNFlaggedSolelyAlgerian{} cited
references to \emph{Journal of Scientific Computing} match this list and no other predatory list, while at least seven backends separately return
legitimate. The list entry appears to refer not to the genuine journal but to a
clone: an archived snapshot of the listed
site presents a journal of that name under a different ISSN (Appendix~\ref{sec:appendix-identity-forensics}).
No name-based matching step can catch that from the citation record
alone, ours included. This is covered by the confidence-weighted vote: those backends outvote the one list.

We did not resolve all \ReliabilityAlgerianMinistryFlags{} matches this
way. We also saw venues listed under their own genuine domain, which points
to a disagreement about criteria rather than a confusion of identities,
and cases of line misalignment in our extraction from the list's PDF. Misalignment can attach a URL to the wrong entry. Among the 20 venues that the list matches most often, it may have affected 2 to 3. We did not count such cases for the whole list.
Dropping this list's vote altogether, rather than down-weighting it,
measurably changes the strict-predatory count.
Section~\ref{sec:leave-one-out} gives the figure and reports it as part of
a range, not as a silent substitution.

\begin{table}[htp]
  \centering
  \footnotesize
  \setlength{\tabcolsep}{4pt}
  \begin{tabular}{@{}lrrr@{}}
    \toprule
    Predatory list & Matches & Leave-focal-out & Second vote \\
    \midrule
    KSCIEN publishers & \ReliabilityKscienPublishersFlags{} & \ReliabilityKscienPublishersLeaveFocalOutPct{} & \ReliabilityKscienPublishersCoFlagPct{} \\
    predatoryjournals.org & \ReliabilityPredatoryjournalsFlags{} & \ReliabilityPredatoryjournalsLeaveFocalOutPct{} & \ReliabilityPredatoryjournalsCoFlagPct{} \\
    Beall's List & \ReliabilityBeallsFlags{} & \ReliabilityBeallsLeaveFocalOutPct{} & \ReliabilityBeallsCoFlagPct{} \\
    KSCIEN standalone journals & \ReliabilityKscienStandaloneJournalsFlags{} & \ReliabilityKscienStandaloneJournalsLeaveFocalOutPct{} & \ReliabilityKscienStandaloneJournalsCoFlagPct{} \\
    Algerian national list & \ReliabilityAlgerianMinistryFlags{} & \ReliabilityAlgerianMinistryLeaveFocalOutPct{} & \ReliabilityAlgerianMinistryCoFlagPct{} \\
    UGC-CARE cloned, Group II & \ReliabilityUgcCareClonedGroupFlags{} & \ReliabilityUgcCareClonedGroupLeaveFocalOutPct{} & \ReliabilityUgcCareClonedGroupCoFlagPct{} \\
    KSCIEN conferences & \ReliabilityKscienPredatoryConferencesFlags{} & \ReliabilityKscienPredatoryConferencesLeaveFocalOutPct{} & \ReliabilityKscienPredatoryConferencesCoFlagPct{} \\
    UGC-CARE delisted, Group II & \ReliabilityUgcCareDelistedGroupFlags{} & \ReliabilityUgcCareDelistedGroupLeaveFocalOutPct{} & \ReliabilityUgcCareDelistedGroupCoFlagPct{} \\
    UGC-CARE cloned, Group I & \ReliabilityUgcCareClonedFlags{} & \ReliabilityUgcCareClonedLeaveFocalOutPct{} & \ReliabilityUgcCareClonedCoFlagPct{} \\
    \bottomrule
  \end{tabular}
  \caption{Corroboration per predatory list. Matches: cited
    references that match the list. Leave-focal-out: percentage of these matches that resolve
    \texttt{predatory} without the list's own vote. Second vote: percentage with
    any other predatory-side vote.}
  \label{tab:reliability}
\end{table}

As a secondary methodology check, we measured pairwise agreement between
the three analyzers across the raw entries of all \PairwiseNNormalFiles{}
normal-sized files (excluding \PairwiseNBloatedFiles{} mega files,
the same $\geq$1{,}000-entry threshold as Section~\ref{sec:bloat}):
OpenAlex and OpenCitations agree on legitimacy \PairwiseOaOcRate\% of
the time they both find the venue (\PairwiseOaOcBothFound{} entries found by both, whether
or not they agreed); Crossref and OpenCitations agree \PairwiseCrOcRate\%
(\PairwiseCrOcBothFound{} entries found by both); OpenAlex and Crossref agree
\PairwiseOaCrRate\% (\PairwiseOaCrBothFound{} entries found by both). These sources may share
underlying indexing coverage, so agreement is not independent
confirmation, and we did not examine why OpenAlex and Crossref
disagree more often.

\subsection{No-citation-log sensitivity}
\label{sec:no-sidecar-sensitivity}
Section~\ref{sec:sibling-inference} sets \CorpusUnverifiedEntries{}
entries aside as unverifiable
(\DerivedUnverifiedPctOfCandidate\% of the candidate population): they
come from the files where neither the file itself nor a sibling yields
a citation log. Nothing shows that these entries are cited at all. To see
whether they could still matter, we folded them back in, each with the
assessment the tool gave it. The identity test of
Section~\ref{sec:identity-misattribution} was applied to them in the same way
as to the primary population, and the decision rule was recomputed without the
invalid votes. Strict-predatory cited
references go from \NosidecarcorrAStrictN{} to \NosidecarcorrCStrictN{},
broad-predatory from \NosidecarcorrABroadN{} to \NosidecarcorrCBroadN{}. The
broad-predatory rate does not rise but falls:
\NosidecarcorrCBroadPct\% instead of \NosidecarcorrABroadPct\%. The
remainder alone has lower rates than the population we analyze:
\NosidecarcorrBStrictPct\% strict and \NosidecarcorrBBroadPct\% broad.

\subsection{\texttt{\textbackslash nocite\{\}} sensitivity}
\label{sec:nocite-sensitivity}
An entry that a \texttt{\textbackslash nocite\{\}} command adds to the bibliography is not cited in the text, so we test whether such entries change the result.

The combined signal (Section~\ref{sec:citation-log}) promotes
\NociteNPromotedTotal{} entries into confirmed-cited status. It uses the
file's compiled bibliography and a second, more complete scan of the \LaTeX{} source. This signal cannot tell a genuine citation apart from one
swept in via \texttt{\textbackslash nocite\{\}}:
\NociteNPromotedAttributable{} of these
\NociteNPromotedTotal{} entries (\NocitePctOfPromotedAttributable{}\%)
trace specifically to \texttt{\textbackslash nocite\{\}}. Of those \NociteNPromotedAttributable{}
entries, \NociteNPromotedAttributableStrict{} (\NocitePctStrictOfAttributable\%) are strict-predatory
and \NociteNPromotedAttributableBroad{} (\NocitePctBroadOfAttributable\%) broad-predatory.

As a sensitivity check, we excluded all \NociteNPromotedAttributable{}
of them outright. This leaves \NociteExcludedStrictCount{} strict and
\NociteExcludedBroadCount{} broad cited references against an
\NociteExcludedDenominator{}-entry denominator. The resulting rates
--- \NociteExcludedStrictRate{}\% and \NociteExcludedBroadRate{}\% ---
are unchanged at the currently reported precision, as is the primary,
resolved-eligible strict rate (\PrimaryStrictRatePct\%, Table~\ref{tab:rates}).

\subsection{Unknown-category sensitivity}
\label{sec:unknown-sensitivity}
About \HeadlineUnknownPct\% of the cited references end up as \texttt{unknown}, so this subsection asks how much that gap could change the strict count.

The headline strict rate covers resolved-eligible entries only.
The \texttt{unknown} cited references are left out, because we do not know how many of them
are predatory. Table~\ref{tab:unknown-sens} therefore makes no assumption about it. It shows what the strict-predatory count would become for a range of hypothetical
predatory rates among the \HeadlineUnknownN{} \texttt{unknown} cited references, using
the \StrictCount{} baseline. These are what-if values, not estimates.

\begin{table}[htp]
  \centering
  \begin{tabular}{@{}rrr@{}}
    \toprule
    Assumed \% predatory & Strict count & Times observed \\
    \midrule
    0 (observed) & \UnknownsensZeroCount{} & \UnknownsensZeroMultiple{} \\
    0.01 & \UnknownsensOnehundredthCount{} & \UnknownsensOnehundredthMultiple{} \\
    0.05 & \UnknownsensFivehundredthsCount{} & \UnknownsensFivehundredthsMultiple{} \\
    0.1 & \UnknownsensOnetenthCount{} & \UnknownsensOnetenthMultiple{} \\
    0.5 & \UnknownsensOnehalfCount{} & \UnknownsensOnehalfMultiple{} \\
    1.0 & \UnknownsensOnepercentCount{} & \UnknownsensOnepercentMultiple{} \\
    \bottomrule
  \end{tabular}
  \caption{Hypothetical strict-predatory count under assumed predatory
    rates among \texttt{unknown} cited references, using the \StrictCount{}
    baseline. Each cell is $\StrictCount{} + \mathrm{round}(\text{rate} \times
    \HeadlineUnknownN{})$, rounded to the nearest integer (not truncated). The last column divides each count by the observed \StrictCount{}.}
  \label{tab:unknown-sens}
\end{table}

Because this pool dwarfs the observed predatory count, even a small
assumed rate moves the headline count by a large multiple: at just 0.1\%
of unknowns assumed predatory, the strict-predatory count would be \UnknownsensOnetenthMultiple{}$\times$ the
observed value (\UnknownsensOnetenthCount{} vs.\ \StrictCount{}); at 1\%, \UnknownsensOnepercentMultiple{}$\times$. 

\subsection{Leave-one-backend-out sensitivity}
\label{sec:leave-one-out}
Table~\ref{tab:reliability} shows that predatory lists disagree with the rest
of the evidence at very different rates. What it does not show is how
much any one list moves the headline numbers. To find that, we drop each
of the 22 backends in turn: we recompute the tool's
confidence-weighted decision rule (Section~\ref{sec:methodology})
without that backend's vote, replaying the rule over the stored results
rather than re-running the tool.

Doing this exhaustively is tractable because the population stays the
same small, fixed set of \BroadCount{} cited references. Two
backends dominate the outcome, and they pull in opposite directions.
Excluding the Algerian national list gives the largest \emph{decrease}
in the strict-predatory count, from \StrictCount{} to
\LeaveOneOutAlgerianMinistryStrictAfterExclusion{}
(\LeaveOneOutAlgerianMinistryStrictPctChange\%). Excluding Scopus gives
the largest \emph{increase}, from \StrictCount{} to
\LeaveOneOutScopusStrictAfterExclusion{}
($+$\LeaveOneOutScopusStrictPctChange\%). The increase is the larger of
the two, and both exceed every other single exclusion.

Across all 22 backends, then, the strict-predatory count runs from
\LeaveOneOutAlgerianMinistryStrictAfterExclusion{} to
\LeaveOneOutScopusStrictAfterExclusion{} depending on which single
backend is left out. The paper's \StrictCount{}
headline sits inside that range rather than at either edge. We read the
range as a robustness check on the estimate, not as evidence that any
one exclusion is the right one to make.

\subsection{Alternative strict definitions}
\label{sec:strict-alternatives}

The strict verdict depends on the tool's confidence weights, so we ask what simpler rules would give.

We also tested two simpler definitions of strict-predatory that count raw
votes without confidence weights. Requiring at least two predatory votes gives
\StrictvoteGeResolvedCorrectedN{} entries,
\DerivedStrictvoteGeResolvedCorrectedMultiple$\times$ the tool's own
\StrictCount{}. Requiring predatory votes to at least equal legitimate votes
gives \StrictvoteVotesGeLegitResolvedCorrectedN{} entries,
\DerivedStrictvoteVotesGeLegitResolvedCorrectedMultiple$\times$
(Table~\ref{tab:strict-alternatives}). Both counts cover the resolved-eligible
population (\ResolvedEligiblePopulationNew{} entries). Neither definition is
stricter than the tool's verdict, which is counter-intuitive. A rule that
sounds like ``needs confirmation from more backends'' is far more permissive.
It admits cases that the weighted formula sends to \texttt{suspicious} or
\texttt{legitimate}, for example single-source cases when predatory votes only
need to at least equal the legitimate votes.

\begin{table}[htp]
  \centering
  \footnotesize
  \begin{tabular}{@{}lrr@{}}
    \toprule
    Definition & Count & vs.\ strict \\
    \midrule
    \textbf{Strict verdict} & \textbf{\StrictCount{}} & --- \\
    At least two predatory votes & \StrictvoteGeResolvedCorrectedN{} & \DerivedStrictvoteGeResolvedCorrectedMultiple$\times$ \\
    Predatory votes $\geq$ legitimate votes & \StrictvoteVotesGeLegitResolvedCorrectedN{} & \DerivedStrictvoteVotesGeLegitResolvedCorrectedMultiple$\times$ \\
    Excl.\ Algerian national list (largest decrease) & \LeaveOneOutAlgerianMinistryStrictAfterExclusion{} & \DerivedAlgerianExclusionStrictMultiple{}$\times$ \\
    Excl.\ Scopus (largest increase) & \LeaveOneOutScopusStrictAfterExclusion{} & \DerivedScopusExclusionStrictMultiple{}$\times$ \\
    \bottomrule
  \end{tabular}
  \caption{Alternative strict definitions, counted over the resolved-eligible
    population.}
  \label{tab:strict-alternatives}
\end{table}

That leaves the two single-backend exclusions of
Section~\ref{sec:leave-one-out}. They show how far the verdict can move
when one source is dropped, but neither exclusion is better founded
than the default, so neither is a candidate to replace it. What this
data supports is a disclosure rather than a replacement.

\subsection{Retraction-history signal}
\label{sec:retraction}
Besides matches with predatory lists, the tool also reports a venue's retraction history, so this subsection shows how common that signal is among the cited references.

Every number in this subsection is a \emph{venue-level, historical}
signal. It records whether the cited \emph{venue} has published
retracted articles at some point in its history. It is not an
\emph{article-level} claim that the cited paper was itself retracted.

Retraction Watch supplies a venue's lifetime retraction records, and
Aletheia-Probe computes a risk tier from them
(\texttt{none}/\texttt{note}/\texttt{low}/\texttt{moderate}/\texttt{high}/\texttt{critical}).
The tool compares a venue's lifetime and recent retraction rates with
thresholds when it knows the venue's publication counts, and its absolute
counts otherwise. One retraction against one recent publication reads as a
100\% recent rate, so we ignore the recent rate for venues with fewer than
\RetractionMinRecentPublications{} recent publications and use the tiers recomputed this way
throughout. In this corpus, \RetractionTinyDenominatorNVenues{} venues change tier through this rule, led by \emph{Journal of Physics
A: Mathematical and General}, which carries
\RetractionTinyDenominatorExampleNCitations{} cited references and an
unremarkable \RetractionTinyDenominatorExampleRatePct\% lifetime rate
across \RetractionTinyDenominatorExampleNPublications{} publications.
The count-based fallback has the opposite problem, for large venues
whose publication volume is not cached: \RetractionCountFallbackNCitations{} cited references across
\RetractionCountFallbackNVenues{} venues, among them BMJ and
\emph{Nature}.

At the venue level, a na\"ive ``ever had any retraction'' signal is
broad. It covers \RetractionNaiveEverRetractedPct\% of all cited
references and says little about a venue's quality on its own. The
tiers separate that population. Of the cited references with any retraction history,
\RetractionNoteTierPctOfAnyHistory\% sit in the lowest \texttt{note}
tier, and \RetractionModerateOrHigherN{} reach \texttt{moderate},
\texttt{high} or \texttt{critical}. Those
\RetractionModerateOrHigherN{} are \RetractionModerateOrHigherPct\% of
all cited references. Of them, \RetractionModerateOrHigherRateBasedN{}
(\RetractionModerateOrHigherRateBasedPct\%) are classified by the
publication-volume-normalized rate and \RetractionModerateOrHigherCountBasedN{}
by the absolute count.

These tiers are what we use wherever retraction risk feeds a statistic
elsewhere in this paper, including the combined ``any documented
concern'' criterion (Section~\ref{sec:headline-prevalence}). The
tool's documentation describes how its tiers are
computed~\citep{aletheia_probe}.

\subsection{Venue concentration}
\label{sec:venue-concentration}
The question here is whether predatory citations are spread across many venues or concentrated on a few. This section compares two populations. One is the \StrictCount{}
strict-predatory cited references, spread over
\ConcStrictDistinctVenues{} distinct venues. The other is the
\ConcLegitTotalCitations{} legitimate cited references, spread over
\ConcLegitDistinctVenues{} venues.

A residue remains on each side. The residue is small, but most
of the strict side could not be identity-checked at all (Section~\ref{sec:identity-misattribution}),
so the measures below are conditional on those matches. On the strict-predatory side the residue is
the \StrictUndecided{} cited references (Appendix~\ref{sec:appendix-corrected-count}) whose
\texttt{predatory} verdict rests on a cloned-journal-list vote that cannot confirm venue identity. We looked up which venues those
belong to: all are already among the
\ConcStrictDistinctVenues{}, so they add no venue of their own. On the
legitimate side the residue is the \LegitreclassifiedN{} cited
references that the identity audit (Section~\ref{sec:identity-misattribution}) moved into \texttt{legitimate}. At \LegitreclassifiedPct\% of
\ConcLegitTotalCitations{} they are too few to change that side's
concentration measures.

Top-$N$ shares alone can mislead when the two populations differ so
much in how many venues they span, roughly 100 times as many on the
legitimate side.
Table~\ref{tab:concentration-measures} therefore reports
several standard measures side by side rather than relying on one, and
its last two rows give each population's own distinct-venue share, not
a share of the corpus. One of those measures is the
Herfindahl--Hirschman index (HHI)~\citepe{rhoades1993herfindahl}{HeinOnline landing page (79 Fed. Res. Bull. 188, 1993): HHI explained as a concentration measure, computed by squaring the market shares of all firms and summing the squares; footnotes name Hirschman (1945) and Herfindahl (1950). Checked against the landing page, not the full text, on 2026-10-07.},
reported together with the effective number of venues, $1/\text{HHI}$,
which is the number of equally-sized venues that would produce the same
concentration.

Four of the measures are \emph{absolute}: top-10 share, top-50 share,
HHI, and the effective number of venues. All four make predatory cited
references the more concentrated. The top 10 strict-predatory venues
hold \ConcStrictToptenPct\% of all \StrictCount{} strict-predatory
cited references and the top 50 hold \ConcStrictTopfiftyPct\%, against
\ConcLegitToptenPct\% and \ConcLegitTopfiftyPct\% on the legitimate
side. The effective number of venues is far smaller as well,
\ConcStrictEffectiveVenues{} against \ConcLegitEffectiveVenues{}.

Two further measures are \emph{relative to each population's own
size}: the Gini coefficient~\citepe{gini1912variabilita}{Scan of the 1912 booklet (Bologna, Tipografia di Paolo Cuppini; extract from Studi economico-giuridici, Cagliari, Anno III, Part 2). Part I introduces the mean difference between observed quantities as a variability index. Section 45, printed p. 80, eq. (112): in the case of maximum inequality (one quantity equals the sum T, all others are zero) the mean difference equals twice the arithmetic mean, so mean difference divided by twice the mean runs from 0 to 1, which is the normalisation of the Gini coefficient. The ratio itself is not written out as a named formula on the pages read (printed pp. 62 to 83, 109); the coefficient is conventionally credited to this work; a footnote points to Gini's earlier Indici di concentrazione e dipendenza. Checked 2026-10-07.}, and the number
of venues needed to reach 50\% or 90\% of cited references as a share
of that population's distinct venues. Both reverse the direction.
Legitimate cited references come out the more unequally distributed, at
a Gini of \ConcLegitGini{} against \ConcStrictGini{}. And half of all
legitimate cited references sit at \ConcLegitVenuesForFiftypctN{}
venues, only \ConcLegitVenuesForFiftypctPct\% of that side's distinct
venues, while half of the strict-predatory ones need
\ConcStrictVenuesForFiftypctN{} venues, or
\ConcStrictVenuesForFiftypctPct\% of the \ConcStrictDistinctVenues{}.

The disagreement is not noise. With roughly 100 times more distinct
legitimate venues to spread across, absolute and population-relative
measures can point different ways, and neither is more correct than the
other. So the claim that survives both is narrower than a flat ``far
more concentrated'': predatory cited references concentrate into a
smaller absolute number of venues. It does not follow that predatory
citation behavior is more unequal in general.

One detail of the ranking matters here: it counts confirmed-cited
entries, not raw entries from a bibliography source. Otherwise
bundled-database venues such as the ACL Anthology would dominate on
sheer uncited volume rather than on actual citation frequency
(Section~\ref{sec:bloat}).

\begin{table}[htp]
  \centering
  \footnotesize
  \begin{tabular}{@{}lrr@{}}
    \toprule
    Measure & Strict-predatory & Legitimate \\
    \midrule
    Distinct venues & \ConcStrictDistinctVenues{} & \ConcLegitDistinctVenues{} \\
    Top-10 share & \ConcStrictToptenPct\% & \ConcLegitToptenPct\% \\
    Top-50 share & \ConcStrictTopfiftyPct\% & \ConcLegitTopfiftyPct\% \\
    HHI & \ConcStrictHhi{} & \ConcLegitHhi{} \\
    Effective venues ($1/\text{HHI}$) & \ConcStrictEffectiveVenues{} & \ConcLegitEffectiveVenues{} \\
    Gini coefficient & \ConcStrictGini{} & \ConcLegitGini{} \\
    Venues for 50\% of cited references & \ConcStrictVenuesForFiftypctN{} (\ConcStrictVenuesForFiftypctPct\%) & \ConcLegitVenuesForFiftypctN{} (\ConcLegitVenuesForFiftypctPct\%) \\
    Venues for 90\% of cited references & \ConcStrictVenuesForNinetypctN{} (\ConcStrictVenuesForNinetypctPct\%) & \ConcLegitVenuesForNinetypctN{} (\ConcLegitVenuesForNinetypctPct\%) \\
    \bottomrule
  \end{tabular}
  \caption{Concentration measures for strict-predatory versus legitimate
    cited references. All strict-predatory figures are computed on the paper's \StrictCount{}-cited-reference
    headline basis, the legitimate figures on \ConcLegitTotalCitations{} legitimate cited references. Effective venues is
    $1/\text{HHI}$ computed from the unrounded HHI, not from the
    3-decimal value shown above it (dividing 1 by the displayed \ConcLegitHhi{}
    would give \DerivedLegitHhiDisplayedReciprocal{}, not \ConcLegitEffectiveVenues{}).}
  \label{tab:concentration-measures}
\end{table}

\subsection{ISSN presence by assessment class}
The identity test relies on the ISSN, so we ask whether strict-predatory references carry one as often as legitimate ones.

Under the strict classification, this paper's primary criterion
(Section~\ref{sec:strict-broad}), legitimate cited references
carry an ISSN more often than strict-predatory ones: \IssnpresenceLegitPct\%
against \IssnpresencePostcorrPct\% (\IssnpresencePostcorrWithIssnN{} of the
\StrictCount{} cited references), a gap of \IssnpresenceGapPostcorr{} points.

Part of that gap is arithmetic, not a difference in citation hygiene. The
identity test (Section~\ref{sec:identity-misattribution}) compares a citing
entry's own ISSN against the ISSN of the predatory-list record that matched it, so
it could only remove cited references that carry an ISSN. Removing
ISSN-carrying references from the strict-predatory population lowers the share
of the remainder that carries one. Whatever true difference exists between
predatory and legitimate citation practice sits underneath that effect.

These figures count confirmed-cited entries only. We did not compute
the same comparison over raw entries.

\begin{widetable}[t]
  \centering
  \footnotesize
  \begin{tabular}{@{}lrrrr@{}}
    \toprule
    Group & Subm.\ (\% of \DiscCorpuswideNSubmissions{}) & Mean entries/subm.\ & Strict \% [95\% CI] & Broad \% [95\% CI] \\
    \midrule
    Computer Science & \DiscCsNSubmissions{} (\DiscCsPctOfCorpus\%) & \DiscCsMeanEntries{} & \textbf{\DiscCsStrictPct{}} [\DiscCsStrictCiLo{}, \DiscCsStrictCiHi{}] & \DiscCsBroadPct{} [\DiscCsBroadCiLo{}, \DiscCsBroadCiHi{}] \\
    Other & \DiscOtherNSubmissions{} (\DiscOtherPctOfCorpus\%) & \DiscOtherMeanEntries{} & \DiscOtherStrictPct{} [\DiscOtherStrictCiLo{}, \DiscOtherStrictCiHi{}] & \DiscOtherBroadPct{} [\DiscOtherBroadCiLo{}, \DiscOtherBroadCiHi{}] \\
    Mathematics & \DiscMathNSubmissions{} (\DiscMathPctOfCorpus\%) & \DiscMathMeanEntries{} & \DiscMathStrictPct{} [\DiscMathStrictCiLo{}, \DiscMathStrictCiHi{}] & \textbf{\DiscMathBroadPct{}} [\DiscMathBroadCiLo{}, \DiscMathBroadCiHi{}] \\
    Physics & \DiscPhysicsNSubmissions{} (\DiscPhysicsPctOfCorpus\%) & \DiscPhysicsMeanEntries{} & \DiscPhysicsStrictPct{} [\DiscPhysicsStrictCiLo{}, \DiscPhysicsStrictCiHi{}] & \DiscPhysicsBroadPct{} [\DiscPhysicsBroadCiLo{}, \DiscPhysicsBroadCiHi{}] \\
    \midrule
    \emph{All} & \DiscCorpuswideNSubmissions{} (\DiscCorpuswidePctOfCorpus\%) & \emph{\DiscCorpuswideMeanEntries{}} & \emph{\DiscCorpuswideStrictPct{}} [\DiscCorpuswideStrictCiLo{}, \DiscCorpuswideStrictCiHi{}] & \emph{\DiscCorpuswideBroadPct{}} [\DiscCorpuswideBroadCiLo{}, \DiscCorpuswideBroadCiHi{}] \\
    \bottomrule
  \end{tabular}
  \caption{Submissions citing at least one strict- or broad-predatory
    reference, as \% of each primary arXiv subject group, with Wilson-score~\citepe{wilson1927probable}{pp. 209 to 210: solves (p0 - p)^2 = lambda^2 p q / n for p, giving the interval with center (p0 + t/2)/(1 + t) and half-width sqrt(p0 q0 t + t^2/4)/(1 + t), t = lambda^2/n (the Wilson score interval); lambda is left general, the 95 percent level corresponds to lambda = 1.96. Checked against the JSTOR scan on 2026-10-07.}
    95\% confidence intervals. Mean entries/subm.\ is mean confirmed-cited
    entries per submission.}
  \label{tab:disciplinary}
\end{widetable}

\subsection{Disciplinary breakdown}
\label{sec:disciplinary-breakdown}
Predatory citation may differ between fields, so we break the rates down by arXiv subject area.

Table~\ref{tab:disciplinary} reports the strict and broad
predatory-citation rate per submission, grouped by primary arXiv
subject category. Section~\ref{sec:disciplinary} discusses what that
pattern does and does not support.

\section{Validation}
\label{sec:validation-power}
We considered adding a manually reviewed, human-judged validation
sample to the tool's own cross-source corroboration, and decided
against it, chiefly because of the sample size needed for false negatives.

Table~\ref{tab:headline}'s final-assessment distribution is heavily
skewed. Against the
\StrictCount{} strict-predatory cited references at
\ConcStrictDistinctVenues{} venues, \texttt{legitimate} holds
\HeadlineLegitimateN{} entries and \texttt{unknown} another
\HeadlineUnknownN{}. Together that is
\DerivedLegitPlusUnknownStrictMultiple$\times$ the strict-predatory
set, or \DerivedUnknownStrictMultiple$\times$ for \texttt{unknown}
alone and \DerivedLegitStrictMultiple$\times$ for \texttt{legitimate} alone.
Their \DerivedLegitPlusUnknownDistinctVenues{} combined distinct venue names
(Section~\ref{sec:venue-concentration} and
Appendix~\ref{sec:appendix-sensitivity}) outnumber the strict-predatory venues
by roughly \DerivedLegitUnknownVenuesPerStrictVenue{}$\times$ (name variants of one
venue count separately).

This is the base-rate effect that makes screening tests for rare
diseases counterintuitive. A false positive (FP), a citation wrongly labeled
\texttt{predatory}, is cheaper to check, because the strict-predatory set is
small. A false negative (FN), one that is actually predatory but labeled
\texttt{legitimate} or \texttt{unknown}, has to be found by sampling
the huge set. Table~\ref{tab:validation-asymmetry} puts numbers on
that. Its Sample $n$ column gives $n\approx 3/p$, the number of
citations needed for a 95\% chance of drawing at least one example of
that row's rate under independent sampling. Citations cluster within
venues in practice, so such a sample covers far fewer distinct venues
than citations.

The asymmetry is severe. An illustrative 10\% false-positive rate inside the
strict-predatory set implies about \DerivedValidationFpTenPctN{} errors. A
false-negative rate of 0.01\%, a thousand times lower, already implies
about \DerivedValidationFnHundredthPctN{} hidden predatory cited
references, spread across a venue space \DerivedLegitUnknownVenuesPerStrictVenue{}$\times$ larger to search.
Showing that the hidden rate is no higher than our own strict rate's
order of magnitude, \PrimaryStrictRatePct\%, would need $n \gtrsim
3/\DerivedPrimaryStrictRatePrinted{} \approx \DerivedRuleOfThreeAtPrintedRateN{}$ manually reviewed cited references with zero
positives found. This is the rule of three, an approximation of the
one-sided 95\% upper Clopper--Pearson bound~\citepe{clopper1934use}{General discussion (p. 404) and the construction of the confidence belt (p. 407): exact confidence limits for a binomial proportion from the binomial terms, confidence coefficients .95 and .99 for central intervals. The rule of three is not in the paper; it approximates the one-sided 95 percent upper limit for zero positives (1 minus 0.05 to the power 1/n, about 3/n). Checked against the JSTOR scan on 2026-10-07.}. A precise point estimate, at 20\%
relative precision and one standard error, would need $n \approx 25/p
\approx \DerivedPrecisionSampleAtPrintedRateN{}$.

Both figures are large for manual review. In a systematic review, the
reviewers timed 73\% of 93 checklists at 0--5 minutes per
use~\citepe{cukier2020checklists}{Abstract and Results (Approximate time for user to complete checklist, p. 5): 68 of 93 checklists (73 percent) completable within five minutes; times measured by the reviewers with a timer emulating a user (Methods), in classes 0-5, 6-10, over 10 minutes. Not times reported by the checklist authors. Checked against the PDF on 2026-10-06.}.
That pace does not scale to a set
\DerivedLegitUnknownVenuesPerStrictVenue{}$\times$ larger.

We did not identify established reliability evidence for human judgments
of predatory status in the sources reviewed here. Across
studies, journal peer review has a mean inter-rater reliability of
$\kappa=0.17$~\citepe{bornmann2010reliability}{Abstract and Table 1 (p. 4): meta-analysis of 70 reliability coefficients from 48 studies; mean Cohen's Kappa .17 (95 percent CI .13 to .21, 26 coefficients); described as 'a slight IRR' after Landis and Koch (p. 4). The label 'none to slight' is not in this paper. Checked against the PDF on 2026-10-06.},
which falls in the ``none to slight'' band of Cohen's scale~\citepe{mchugh2012kappa}{Section on Cohen's kappa (p. 279): Cohen suggested 0.01-0.20 as none to slight, 0.21-0.40 as fair, and so on. McHugh calls this too lenient; her own Table 3 reads 0-.20 as None (0-4 percent of data reliable). Checked against the PDF on 2026-10-06.}. \citete{cukier2020checklists}{The review reports no reliability figure for any checklist; it assesses risk of bias only (five questions, Methods; Table 2). Only 3 of 93 are rated evidence-based. Checked against the PDF on 2026-10-06.} report no reliability figure for any of the 93
checklists they review.
\citete{nghaynes2021checklists}{Abstract (p. 2): four evidence-based checklists identified (five risk-of-bias questions of Cukier et al.); none completed the first of nine steps of scale development; conclusion: lack assessment of reliability and validity. Results (p. 14): no step beyond 1.5, including tests of reliability and validity. Checked against the postprint on 2026-10-06; volume and pages not in the postprint.} found that none of four checklists rated evidence-based had been assessed for reliability or validity.
And in one study of physiotherapy trials, inter-rater reliability for rating
study quality was lower for predatory-journal
publications~\citepe{paci2022pedro}{Abstract and Discussion (p. 6): 298 RCTs from 18 physiotherapy journals (9 on Beall's list, others in DOAJ), 2014-2017, rated with the PEDro scale. ICC .537 (PJs) against .729 (NPJs); kappa for moderate-to-high against low quality .28 against .57. Raters not blinded to journal source; criteria for predatory not fully defined (Limitations). Checked against the PDF on 2026-10-06.}.

These figures suggest, by analogy, that holistic quality judgments can
be unreliable. They do not measure agreement on predatory status, and
they say nothing about the mechanical checks used here.

So a random citation-level sample large enough to pin down corpus-wide
false-negative prevalence would be a large effort, and we make no claim to
estimate recall against human ground truth. Human judgment would not
obviously be the more stable target anyway, since curated predatory lists
already differ in how often other sources corroborate them
(Section~\ref{sec:backends}).

Deciding whether an individually matched venue really is predatory
therefore lies outside this paper's design, because it would require an
independent quality signal we do not have. What the paper checks
instead is narrower: whether a predatory list's match points at the entity
it claims to (Section~\ref{sec:identity-misattribution}), how far
the lists agree with each other
(Section~\ref{sec:backends}), and how sensitive the verdict is to
any single list (Section~\ref{sec:leave-one-out}). Checking
against already-labeled external lists avoids the sample-size problem,
because the problematic venues are identified in advance.
Appendix~\ref{sec:appendix-benchmark} reports such checks against two
lists of problematic venues curated by third parties. We do not treat
these lists as gold-standard labels for predatory status.

\begin{table}[htp]
  \centering
  \footnotesize
  \begin{tabular}{@{}lrrrrr@{}}
    \toprule
    Bucket & Venues & Rate & Errors & vs.\ FP & Sample $n$ \\
    \midrule
    Pred & \ConcStrictDistinctVenues{} & 10\% FP & \DerivedValidationFpTenPctN{} & \DerivedValidationFpTenPctMultiple{}$\times$ & 30 \\
    \midrule
    L+U & \DerivedLegitPlusUnknownDistinctVenues{} &0.001\% FN & \DerivedValidationFnThousandthPctN{} & \DerivedValidationFnThousandthPctMultiple{}$\times$ & 300{,}000 \\
    L+U & \DerivedLegitPlusUnknownDistinctVenues{} &0.01\% FN & \DerivedValidationFnHundredthPctN{} & \DerivedValidationFnHundredthPctMultiple{}$\times$ & 30{,}000 \\
    L+U & \DerivedLegitPlusUnknownDistinctVenues{} &0.1\% FN & \DerivedValidationFnTenthPctN{} & \DerivedValidationFnTenthPctMultiple{}$\times$ & 3{,}000 \\
    L+U & \DerivedLegitPlusUnknownDistinctVenues{} &1\% FN & \DerivedValidationFnOnePctN{} & \DerivedValidationFnOnePctMultiple{}$\times$ & 300 \\
    \bottomrule
  \end{tabular}
  \caption{Illustrative error counts and review sample sizes by bucket, as
    derived in the text. Venues is the number of distinct venue names in
    the bucket; Errors and Sample $n$ count cited references (\StrictCount{} strict-predatory; \ConcLegitTotalCitations{} legitimate and \HeadlineUnknownN{} unknown) (Pred =
    strict-predatory; L+U = legitimate+unknown; FP/FN as defined in
    the text; vs.\ FP: errors relative to the Pred row).}
  \label{tab:validation-asymmetry}
\end{table}

\subsection{Add-one impact of three candidate evidence sources}
\label{sec:add-one}

This study identified three evidence sources that are not part of
Aletheia-Probe's 22-backend ensemble, and none of them feeds its
confidence-weighted vote today. The first is the ACL Anthology, usable
as a legitimate list. None of its venues is classified as
\texttt{predatory} or \texttt{hijacked} by the existing backends
(Section~\ref{sec:bloat} gives the numbers). The
second is the Retraction Watch Hijacked Journal
Checker~\citepe{retraction_watch_hijacked_checker}{Main page and Methods page: cases found or confirmed by comparing journal archives, website templates, growth in indexing and anomalous citations, plus checks of title, ISSN, indexation, domain registration, editorial board and archives; database covers hijacked journals located or active from 2020 on; the Original Journal column can be uncertain. The wording 'forensically verified' is not on these pages. Pasted by the author on 2026-10-07.}, which contributes
\HijackedbenchRwTotal{} independently curated cases, found by analyzing journal archives, website templates and indexing patterns. The third is DOAJ's own change log of
journals withdrawn for not adhering to its best practice, with
\HijackedbenchDoajTotal{} cases (the entries with that reason).

How much of that would be new evidence? Matching by ISSN first and by
name as a fallback, only \AddoneRwCoveragePct\% of the Hijacked Journal
Checker's cases and \AddoneDoajCoveragePct\% of the DOAJ cases already
have a matching entry among the tool's nine predatory lists.
Most of the DOAJ cases would therefore be new evidence, and so would
about half of the Hijacked Journal Checker cases. The ACL Anthology
looks the same from the legitimate side:
\AddoneAclPctUnknownSuspicious\% of the corpus citations that match one
of its known venues carry no confident legitimate verdict today, being
either \texttt{unknown} or \texttt{suspicious}.

We simulated adding each source as a new backend, one at a time, with
the same weight as every existing backend. For this simulation we treat
the Hijacked Journal Checker and DOAJ's withdrawal log as predatory-side
evidence. That is a hypothetical mapping. The tool itself keeps its
hijacked-journal list out of the predatory counts (Section~\ref{sec:methodology}),
so the resulting counts are not the strict and broad prevalence defined there. The evaluation run used one
uniform weight for all of its 20 vote-eligible backends
(Section~\ref{sec:methodology}), so the value of that weight does not
change the outcome. Table~\ref{tab:addone} gives the result.

\begin{table}[htp]
  \centering
  \footnotesize
  \setlength{\tabcolsep}{4.5pt}
  \begin{tabular}{@{}lrrrr@{}}
    \toprule
    Source & Affected & Strict & Broad & Conflicts \\
    \midrule
    ACL Anthology & \AddoneAclNAffected{} & \parbox[t]{1.15cm}{\raggedleft \StrictCount{}\\(\AddoneAclStrictDelta{})} & \parbox[t]{1.15cm}{\raggedleft \BroadCount{}\\(0)} & \AddoneAclNConflicts{} \\
    Hijacked Journal Checker & \AddoneRwNAffected{} & \parbox[t]{1.15cm}{\raggedleft \DerivedAddoneRwStrictTotal{}\\($+$\AddoneRwStrictDelta{})} & \parbox[t]{1.15cm}{\raggedleft \DerivedAddoneRwBroadTotal{}\\($+$\AddoneRwBroadDelta{})} & \AddoneRwNConflicts{} \\
    DOAJ withdrawal & \AddoneDoajNAffected{} & \parbox[t]{1.15cm}{\raggedleft \DerivedAddoneDoajStrictTotal{}\\($+$\AddoneDoajStrictDelta{})} & \parbox[t]{1.15cm}{\raggedleft \DerivedAddoneDoajBroadTotal{}\\($+$\AddoneDoajBroadDelta{})} & \AddoneDoajNConflicts{} \\
    \midrule
    \emph{Combined} & \emph{\AddoneCombinedNAffected{}} & \parbox[t]{1.15cm}{\raggedleft\emph{\DerivedAddoneCombinedStrictTotal{}}\\\emph{($+$\AddoneCombinedStrictDelta{})}} & \parbox[t]{1.15cm}{\raggedleft\emph{\DerivedAddoneCombinedBroadTotal{}}\\\emph{($+$\AddoneCombinedBroadDelta{})}} & \emph{---} \\
    \bottomrule
  \end{tabular}
  \caption{Add-one impact of the three candidate evidence sources.
    Affected is the number of cited references with a match in the
    source (ISSN first, then name). Strict and Broad give the resulting
    counts, with the change from the \StrictCount{}/\BroadCount{}
    baseline in parentheses, and Conflicts override an existing
    \texttt{legitimate} verdict.
    \emph{Combined} is not the row sum: the \AddoneCombinedNMultiSource{} cited references that
    match more than one source are counted once, but they receive the votes of both
    predatory-side sources, which can turn a \texttt{legitimate} verdict into \texttt{predatory} when
    neither source alone does.}
  \label{tab:addone}
\end{table}

Adding the ACL Anthology as a legitimate list moves
\AddoneAclMovedToLegitimateN{} cited references from \texttt{unknown}
to \texttt{legitimate} and \AddoneAclMovedSuspiciousToLegitimateN{}
from \texttt{suspicious}, with no effect at all on either headline
predatory count, because none of the affected citations carried
predatory-side evidence to begin with. The strict-predatory count moves
to \DerivedAddoneRwStrictTotal{} with the Hijacked Journal Checker
added, \DerivedAddoneDoajStrictTotal{} with DOAJ's withdrawal log, and
\DerivedAddoneCombinedStrictTotal{} with all three together. The
broad-predatory count moves to \DerivedAddoneRwBroadTotal{},
\DerivedAddoneDoajBroadTotal{} and \DerivedAddoneCombinedBroadTotal{}
respectively.

For the two predatory-side sources, every movement runs in one
direction. No cited reference loses a \texttt{predatory} verdict when
one of them is added; it either gains one or keeps the one it had (\AddoneRwNPredatoryKept{} cases for the Hijacked
Journal Checker, \AddoneDoajNPredatoryKept{} for DOAJ). That is a structural property of the
vote rather than a finding about these two sources, since adding
predatory-side evidence can only push the aggregate score further
toward \texttt{predatory}.

The table's \emph{Conflicts} column runs the other way. Those are cases
where an existing backend had already resolved the citation
\texttt{legitimate} and the new source's evidence is strong enough to
override it. A conflict invites the reading that the tool had it
wrong and the new source caught something. For DOAJ that reading is
uncertain. \DoajReinclusionPctReinstated\% of all checkable withdrawn journals are
currently reinstated on DOAJ's own live listing, and among
historically withdrawn journals that resolve \texttt{legitimate} today,
\DoajReinclusionLegitresolvingPctStillRemoved\% remain absent from that
listing regardless. Being withdrawn from DOAJ, and staying withdrawn,
is not equivalent to a predatory label
(Appendix~\ref{sec:appendix-benchmark}).

None of this argues that either source should be added to the tool as it
stands, nor that the current \StrictCount{}/\BroadCount{} headline is
wrong. It shows how far the estimate moves if evidence coverage is
extended in this specific, unweighted way, one source at a time, in the
same spirit as the leave-one-backend-out analysis running the other direction
(Section~\ref{sec:leave-one-out}).
Appendix~\ref{sec:appendix-benchmark} asks the reverse question,
whether the tool as it stands already recognizes these same venues.

\subsection{Identity misattribution in name-based venue matching}
\label{sec:identity-misattribution}

This section runs a mechanical identity check. For each originally
strict-predatory cited reference that carries its own ISSN, we compare
that ISSN against the ISSN of whichever predatory-list record won the
citation's \texttt{predatory} vote. Across all
\AuditOriginalStrictTotal{} originally strict-predatory cited
references, the test finds \AuditMismatch{} mismatches
(\DerivedAuditMismatchPctOfOriginal\%). It cannot check every case:
\AuditNoIssn{} of them (\AuditNoIssnPctOfRaw\%) carry no ISSN of their
own, so there is nothing to compare against. That is the measured coverage limit of the ISSN check
(Section~\ref{sec:limitations}).
Appendix~\ref{sec:appendix-identity-forensics} gives the matching mechanism
and the case of \emph{Journal of Scientific Computing}.

The corrected headline counts used throughout this paper come from a
full replay of the confidence-weighted decision rule with each
confirmed-cited entry's invalid vote removed, not from subtracting all
\AuditMismatch{} outright. Strict-predatory goes from
\AuditOriginalStrictTotal{} to \StrictCount{}
($-$\DerivedCorrectionSizePct\%), and broad-predatory from \CorrectionBeforeBroadCitationsN{} to \BroadCount{}
by the same replay. The replay reaches only the references that the audit tested, the strict ones.

The corrected \StrictCount{} then partitions four ways, summing to
\StrictCount{} exactly. \StrictIdentityVerified{} are
identity-verified. \StrictIdentityUnverified{} are
identity-unverified, meaning an ISSN match that the winning predatory list never verified. \StrictNoIssn{} (\StrictNoIssnPct\%) carry
no citing-side ISSN at all. And \StrictUndecided{} are
clone-undecidable, resolvable from neither.
Appendix~\ref{sec:appendix-identity-forensics} gives the derivation of the corrected count and of this partition.

\section{Discussion}
\label{sec:discussion}

\textbf{Headline rate versus prior estimates.}
The number a reader most likely wants is the fraction of cited references in the
covered submissions that point at a predatory journal. At the precision this
study can support, that number has no single correct value. We report
several, under different denominators, rather than collapsing them into
one (Table~\ref{tab:rates}), and use the strict figure as the primary
operational estimate. The reason is Section~\ref{sec:backends}'s finding
that the broad number is heavily driven by one list whose matches the
rest of the tool does not corroborate in
\DerivedAlgerianMinistryNotRetainedPct\% of cases. Corroboration is not
validation, so this is a choice of default and not evidence that strict is
closer to the true value.

Both numbers are small in absolute terms. They measure cited references,
not published articles, so they are not directly comparable to estimates of predatory-journal
\emph{article} volume, such as the estimated 420{,}000 articles in
2014~\citepe{shen2015predatory}{Abstract (p. 1): estimated 420,000 articles in 2014 from a stratified sample, described as rough in the study's limitations. Fig. 4 (p. 8): article volume of predatory journals by discipline in absolute numbers for 2014 (general about 162,000, engineering 97,000, biomedicine about 70,000). The article gives no field-level shares of all articles. Checked against the PDF on 2026-10-06.}. We expected arXiv's
own selection effect to explain a low rate, since its dominant
disciplines are computer science, physics and mathematics rather than
some of the disciplines more strongly represented in the prior work of
Section~\ref{sec:related}.  We also
cannot rule out a methodological undercount, given \texttt{unknown}'s
\HeadlineUnknownPct\% share of confirmed-cited entries
(Section~\ref{sec:headline-prevalence}).

\label{sec:disciplinary}
\textbf{Disciplinary composition, partly tested.}
Table~\ref{tab:disciplinary} (Section~\ref{sec:disciplinary-breakdown})
neither supports nor refutes that explanation, because it compares only
disciplines within the corpus.
At the submission level, Computer Science has the highest
strict-predatory rate of any group, \DiscCsStrictPct\%, and it was one
of the three disciplines we expected to be lower-risk. That lead coincides with longer bibliographies than in Other and
Mathematics (Table~\ref{tab:disciplinary}'s Mean entries/subm.\ column);
Physics has the longest and a lower rate, so length alone does not
explain the order. At the
citation level, Computer Science and Other are nearly tied
(\DiscciteCsStrictPct\% and \DiscciteOtherStrictPct\%, against a
\MultibibCurrentStrictRate\% baseline), and both stay well above
Mathematics and Physics (\DiscciteMathStrictPct\% and \DisccitePhysicsStrictPct\%).

This analysis cannot
settle the wider question, because arXiv's ``Other'' bucket consists
mostly of quantitative and technical categories. Whether a corpus spanning medicine or the
humanities, where much of the prior work sits, would
show the effect the disciplinary argument invoked is beyond what this
corpus can say.

Mathematics is the one exception, with a submission-level and
citation-level broad rate well above baseline. At the citation level it is not a
disciplinary signal. \MathjscPctJsc\% of its broad-predatory cited
references trace to one venue, \emph{Journal of Scientific Computing}
(Appendix~\ref{sec:appendix-identity-forensics}), which the tool's
own full-evidence assessment classifies \texttt{legitimate}. Exclude
it and Mathematics's citation-level broad rate falls below the corpus
baseline. Appendix~\ref{sec:appendix-disciplinary} gives the full
derivation.

\textbf{Bibliography bloat and citation logs.}
Of everything reported here, the bibliography-bloat finding
(Section~\ref{sec:bloat}) is in our view the most broadly actionable. A citation log built from the \LaTeX{} source alone is not enough,
because it misses keys that custom citation commands hide. The keys that
the combined signal recovers are therefore added
(Section~\ref{sec:nocite-bbl}). We recommend both as sanity
checks for future large-BibTeX-corpus studies.

\textbf{Low-cost operational uses.}
A low-cost use of an automated check of this kind could be explored. Individual authors
could run such a tool on their own bibliography before submitting, and check each matched reference by hand, for example by comparing the ISSN of the matched predatory-list entry with the ISSN of the journal they meant to cite. A warning shown only to the
submitting author is a different harm profile
from a public, verdict-linked disclosure of individual cases. We ran the tool on this paper's own reference list. None of the entries that the tool assesses, journal articles and conference papers, matches a predatory list. The tool does not assess the other entries, which are web pages, software, books, book chapters and preprints.

\textbf{The missing venue identifier.}
One broader observation this work keeps running into: academic
publishing lacks, as far as we are aware, a canonical, tamper-resistant
venue identifier the way DOI identifies a paper or ORCID an author. ISSN comes closest, but a
citing entry can claim any ISSN without that claim being checked
against the authoritative ISSN Register, and a venue can lack one
altogether. That may be one reason why the \emph{Journal of Scientific Computing}
clone could impersonate a real journal under a different ISSN, and why
\HeadlineUnknownPct\% of confirmed-cited entries remain unresolved by
the assessment backends (Section~\ref{sec:headline-prevalence}). A registry-backed
identifier would not fix predatory publishing on its own, but it could
remove part of the ambiguity that measurements like this one work
around.

\section{Limitations}
\label{sec:limitations}

\textbf{Corpus scope.} Findings characterize venues cited by the covered submissions, not citation practices in general. arXiv's
own disciplinary composition is dominated by computer science, physics
and mathematics, and is not representative of academia as a whole.
Coverage is
also limited to submissions with a \LaTeX{} source package that
produced an assessable \texttt{.bib} file (Section~\ref{sec:corpus}).
The excluded \CorpusExcludedSubmissionsPct\% are not
necessarily representative of arXiv either, and could skew the figures
in either direction. Repositories from other disciplines, such as the life sciences or the humanities, could widen the picture. We did not test any.

\textbf{No country or affiliation stratification.} Two of the studies discussed above, \citet{savina2021prevalence} and
\citete{machacek2022predatory}{Abstract; Introduction: names of potential, possible or probable predatory journals and publishers on Beall's lists mapped to ISSNs (3,293 journals, via Ulrichsweb) and searched in Scopus; 324 journals found with 164,000 articles in 2015 to 2017; share of articles per country for 172 countries in four fields. Counts articles by publishing country, not citing documents. Checked against the PDF on 2026-10-06.}, report predatory-publishing exposure by
country, in journal- and country-level analyses.
This paper does not. The pipeline operates on the citing submission's
own \texttt{.bib} file, and author affiliation and country are neither
an extraction target nor consistently structured in arXiv's \LaTeX{}
source. 

\textbf{No within-submission citation deduplication.}
Submissions that bundle the same citation across more than one
\texttt{.bib} file produce \CorpusMultiBibExcessDupRows{} excess
confirmed-cited rows, and this paper does not collapse them. Checked
against the headline populations, \MultibibStrictNExcessRows{} of the
\StrictCount{} strict-predatory cited references
(\MultibibStrictPctOfPopulation\%) and \MultibibBroadNExcessRows{} of
the \BroadCount{} broad-predatory ones
(\MultibibBroadPctOfPopulation\%) are themselves such excess rows.
Collapsing them would move both citation-level rates by under 0.0001
percentage points, and slightly upward, since the headline populations
duplicate less than the corpus-wide average. Every submission-level rate is unaffected.

\textbf{Coverage denominator.} \HeadlineUnknownN{} of the
\EligiblePopulation{} eligible cited references
(\CoverageUnknownOfEligiblePct\%) come out \texttt{unknown}, meaning
unresolved, rather than legitimate or predatory. The backends'
combined coverage, broad as 22 backends are, does not resolve about a
third of the eligible cited references in this corpus.
Section~\ref{sec:denominators} explains why the strict estimate
excludes this population from its denominator while the broad estimate
cannot, and Section~\ref{sec:unknown-sensitivity} quantifies how far it
could move the headline numbers under explicit, disclosed assumptions.

\textbf{Inferring citation status from another file is not a direct
citation-log match.} Section~\ref{sec:sibling-inference} excludes
\CorpusExcludedEntries{} entries from ``actually cited'' by inference:
a file that yields no citation log, but shares an arXiv submission with
a file that does, is taken to be the unused half of a bloated pair.
That is one step weaker  than a direct per-entry citation-log
confirmation, and we say so wherever the excluded population appears.
Section~\ref{sec:sibling-inference} describes the inference and a
consistency check of it. A
further \CorpusUnverifiedEntries{} entries
(\DerivedUnverifiedPctOfCandidate\% of the original candidate
population) have no other file's log covering them either. Those are
excluded from the primary population entirely and reported separately
in Section~\ref{sec:no-sidecar-sensitivity}.

\textbf{Predatory-list cross-source corroboration is heterogeneous,
possibly time-varying, and not a validity measure.} List-level corroboration
with the rest of the tool differs widely. This
may partly reflect each list's own institutional purpose and update cadence
(Section~\ref{sec:backends}, Table~\ref{tab:reliability}). That
heterogeneity is part of why we did not pursue a random-sample manual
validation (Section~\ref{sec:validation-power}) and checked recognition
against two externally curated lists
instead. Frozen predatory-list snapshots can also go stale relative to a
fast-moving upstream source. This paper measures that heterogeneity
rather than adjudicating any individual venue's history against it.

\textbf{Construct validity of the \texttt{legitimate} label.}
This label mostly reflects whether a venue is indexed by
a source such as Crossref, OpenAlex, Scopus, or DBLP. Being indexed
shows a venue is discoverable, not that it has good editorial
practices. So a venue with poor editorial standards can still be
indexed, and this tool would still call it \texttt{legitimate};
the strict-predatory count may therefore \emph{underestimate} true prevalence.
We stress-tested this label with the external benchmark in
Appendix~\ref{sec:appendix-benchmark}: among journals that DOAJ removed for not meeting its
best-practice standards and that no predatory list matches, \DoajxtabLegitPctOfDisjoint{}\% are
still labeled \texttt{legitimate} by this tool. Removal by DOAJ is broader than
being predatory, so this stress-tests the label and is not a direct
estimate of false negatives. The legitimate label is not only due to DOAJ's later
readmissions: \DoajReinclusionLegitresolvingPctStillRemoved\% of these journals are still absent from DOAJ's
current listing. Lists that disagree on a journal are not unique to our setting.
\citete{strinzel2019blacklists}{Abstract and Results (p. 4): 72 journals and 42 publishers appear in both a blacklist and a whitelist (Beall, Cabell blacklist; DOAJ, Cabell whitelist; lists as of 13 Dec 2018); DOAJ and Beall 41 journals (0.3 percent), DOAJ and Cabell blacklist 37 (0.2 percent). Discussion (p. 13): 'Although the overlap was small'. Part of the Cabell overlap reportedly came from a system error (Limitations, p. 13). Checked against the PDF on 2026-10-06.} found a small number of journals that appear on both a
blacklist and a whitelist. We do not correct for this,
since doing so would require an independent quality judgment this
paper does not have available.

\textbf{Missing identifiers and name-only matching drive some of this
ambiguity.} Matching by venue name alone cannot by
itself distinguish a legitimate venue from an unrelated clone sharing
its name, as the \emph{Journal of Scientific Computing} case
illustrates (Appendix~\ref{sec:appendix-identity-forensics}). Missing
identifiers limit what our audit can check. We did not
test how many cases consistent identifier
practice would resolve.

The ISSN comparison cannot check every reference. \StrictNoIssnPct\%
(\StrictNoIssn{} of \StrictCount{}) of the strict-predatory cited references
carry no ISSN of their own (Section~\ref{sec:identity-misattribution}), so most of them could not be checked.
The limit comes from the citing entries, which often carry no ISSN. The same holds on the legitimate side. Of the
\MirrorexhaustiveNTrueCandidates{} cited references labeled legitimate on the strength of the CORE rankings alone, which never record an ISSN and therefore match by name only, \MirrorexhaustivePctNoCitingIssn\% have no ISSN.
Among the \MirrorexhaustiveNCheckable{} that could be checked,
\MirrorexhaustivePctConfirmedMisattribution\% show an ISSN mismatch, against
\AuditMismatchRateOfCheckable\% (\AuditMismatch{} of \AuditCheckableWithIssn{}) on the strict-predatory side. Nearly all of
them are one legitimate venue under several name variants. We found no questionable
venue labeled legitimate, and no count changes.

Most of the \AuditMismatch{} mismatches rest on an inference. Only \SisfIssnlNRefsCacheIssnDiffers{} meet a cache record with a different ISSN. The other \SisfIssnlNRefsNullCacheOnly{} meet a cache record without an ISSN, and the citing ISSN is found on another record. The identity correction also overshoots slightly. The test compares ISSN strings, so the print and the online ISSN of one journal count as a mismatch. Merging such pairs would restore \SisfIssnlNRestored{} references. Besides, the lookup of the corroborating record used a cache that was extended after the evaluation run, which matters for \SisfIssnlNLaterCacheOnly{} references. Together these would raise the strict count from \StrictCount{} to at most \SisfIssnlStrictWithBothN{} (\DerivedSisfIssnlStrictMaxMovementPct{}\%) and the broad count from \BroadCount{} to \SisfIssnlBroadWithBothN{}.

\textbf{Construct validity of the retraction-history signal: its
direction is not settled.} Retracting an article requires a venue to
run a correction process at all; a venue that runs none leaves no
retraction record, whatever its editorial practices. Among 17 journals,
retraction frequency rose with impact
factor~\citepe{fang2011retraction}{Section 'Journals differ in retraction frequency' (p. 3856): PubMed search for retracted articles among 17 journals with impact factors from 2.00 to 53.484; strong Spearman correlation between retraction index and impact factor (P < 0.0001, Fig. 1). p. 3857: higher visibility may attract greater scrutiny, named as one of several possible explanations. Checked against the PDF on 2026-10-07.}, and the authors name the greater
scrutiny of visible journals as one possible reason. Of the 1{,}511
standalone journals on Beall's List, only 46 had retracted anything as
of 2022~\citepe{xu2026retraction}{Abstract: 1,511 standalone potential predatory journals on the updated Beall's List; according to the Retraction Watch Database only 46 had retracted publications (645 in total) as of 2022, 18 of them indexed in the Web of Science Core Collection. Checked against the accepted manuscript (preprint) on 2026-10-07; journal pages from the publisher DOI record.}. A retraction record is therefore in
part a marker of a correction process that works. But mass retractions
after paper-mill detection attach the same signal to venues under acute
quality pressure. Hindawi, for example, retracted more than 8,000
paper-mill articles in 2023~\citepe{kincaid2023hindawi}{Retraction Watch, 2023-12-19: 'So far in 2023, Hindawi has retracted over 8,000 articles'; the post ties the retractions to paper-mill manipulation of special issues. News source relying on the publisher's figure. Page fetched on 2026-10-07.}. So the direction
is ambiguous rather than inverted. The strict and broad figures do not
inherit this, since Retraction Watch contributes to neither count
(Section~\ref{sec:strict-broad}); the
``any documented concern'' criterion of
Section~\ref{sec:headline-prevalence}, which retraction risk drives,
does. We do not test the direction in our own data.

\textbf{Local database snapshots age.} The OpenAlex, Crossref,
OpenCitations and DOAJ mirrors reflect a snapshot as of when they were
built, not live upstream state. That is deliberate, for reproducibility
and to avoid rate limits, but it means very recent venue status changes
are not reflected.

\textbf{Clone pages count both sides of a pair as legitimate.} In the tool
state evaluated here, the UGC-CARE clone pages give each cloned journal a
legitimate vote as well as a predatory one. This holds for all
\UccdvNWithClonedVote{} cited references with a cloned-list vote. Removing the
legitimate vote for exactly these references would move
\UccdvNVerdictChanged{} of them from legitimate to predatory, so the strict
count would be \UccdvStrictAfterN{} instead of \UccdvStrictBeforeN{}. The broad
count does not change. This lies inside the leave-one-backend-out range
(Section~\ref{sec:leave-one-out}).

\textbf{Entry-type parsing noise.} A long tail of malformed
\texttt{.bib} files produces spurious ``entry types'' under
relaxed-mode parsing. We do not name these
individually below a fixed threshold, and did not attempt to recover
their true type. That naming threshold does not affect the eligibility
partition itself, which remains unfiltered.

\textbf{Numerical accuracy is enforced by the build process where practical, not
everywhere.} Most data-driven figures in this paper are emitted as
\LaTeX{} macros by the scripts that compute them, and checked for
cross-script consistency before every build. That is specifically to
avoid the transcription and staleness errors that hand-copying a number
into the text invites. It could not be done for every figure: some
derived quantities and one-off values were typed by hand, from a script's output or from a cited source. The build process reduces the chance that a reported number
is stale or mistyped, but it does not eliminate it.

\section{Conclusion}
\label{sec:conclusion}

This paper had two aims: to measure how often cited references point to venues on predatory lists, and to show how far these numbers can be trusted. Most of the analysis serves the second aim.

We ran the 22-backend tool Aletheia-Probe over the raw
BibTeX entries of the \CorpusCoveredSubmissions{} arXiv submissions this study covers
and restricted the citation analyses to confirmed-cited entries. The submissions are \CorpusCoveredSubmissionsPct\% of the
\CorpusTotalSubmissions{} submissions arXiv received in
2025~\citep{arxiv_monthly_stats}. The tool's strict verdict labels
\PrimaryStrictRatePct\% of cited references with a confident venue match as
predatory. At the submission level, \SubmissionStrictPct\% of the
submissions with a confirmed-cited entry cite at least one such reference.

Both strict rates are operational pipeline estimates, not validated bounds on true
prevalence. Limited predatory-list coverage and missed venues could push them down, while
false-positive identity and classification errors could push them up, and
the present study cannot determine their net effect
(Section~\ref{sec:validation-power}). 

Quantified exposure exists on both sides. One analysis points to a
possible undercount. In the add-one evaluation of three candidate
evidence sources that are not yet in the tool's ensemble, the count
rises from \StrictCount{} to \DerivedAddoneCombinedStrictTotal{} when all three are added, two of
them treated, for the test only, as predatory evidence. That is not the strict estimate
defined in this paper. 

On the overcount side, the identity-misattribution partition is the
structural counterpart. It measures how much of the count rests on
unverified identity, not how many references are wrong. Of the \StrictCount{} strict-predatory cited
references, \StrictIdentityUnverified{} remain identity-unverified,
\StrictNoIssn{} carry no citing-side ISSN at all, and
\StrictUndecided{} are clone-undecidable between a genuine venue and an
impersonator using its stolen ISSN
(Section~\ref{sec:identity-misattribution}). That sits on top of the
order-of-magnitude heterogeneity in predatory-list cross-source
corroboration documented above (Section~\ref{sec:backends}).

Further identity checks could remove more of the
\StrictCount{} cited references that retain the tool's strict verdict and more of the \BroadCount{} broad matches
(Section~\ref{sec:identity-misattribution}).

The gap to the broad estimate (\BroadCount{} against \StrictCount{} cited references) is driven largely by matches from one list
that the rest of the tool does not corroborate. The three methodological findings
may matter beyond this study: the lists disagree with each other
(Section~\ref{sec:backends}), matching venues by name alone can misattribute evidence
(Section~\ref{sec:identity-misattribution}), and raw BibTeX counts overstate real reference lists
(Section~\ref{sec:bloat}). Other large-scale studies on raw \texttt{.bib} files or on name-matched
venues may run into the same traps.

\section{Data Availability}
\label{sec:data-availability}
The data and software that this study builds on are publicly available in their current versions: arXiv's source
packages, the bibliometric databases and predatory-journal lists
listed in Section~\ref{sec:methodology}, and
the open-source Aletheia-Probe
tool~\citepe{aletheia_probe,florath2026aletheia}{Abstract: open-source tool, repository https://github.com/sustainet-guardian/aletheia-probe. Checked against the submission version on 2026-10-06.}.  The
scripts that build the corpus and compute this paper's results are
released separately, as code only~\citep{predatory_citations_arxiv_2025_repo}.

We hold back two things on purpose. The first is any list of
venues, predatory or legitimate. Because of the identity gaps described
in Section~\ref{sec:identity-misattribution}, such a list would suggest
a precision this measurement cannot deliver.  The second is the pipeline's per-citation
output. It would link a named arXiv submission to a specific predatory,
legitimate or hijacked verdict, including the source conflicts and
false-positive examples documented in this study. We consider
that inappropriate, because it would turn an aggregate finding into a
signal about individual authors. We do not redistribute the bibliography entries taken from a paper's \LaTeX{} source either.
arXiv's catalog metadata is CC0, but its e-print full text is restricted, and we release only aggregate results.

Every intermediate artifact, meaning the extracted \texttt{.bib}
entries, citation logs, per-backend query results and venue-level
aggregates, comes from running arXiv's public source packages through
the tool against the locally hosted snapshots of the evidence sources
listed in Section~\ref{sec:methodology}. 

\section{Conflict of Interest}
\label{sec:coi}
The author is also the developer of
Aletheia-Probe~\citepe{aletheia_probe,florath2026aletheia}{Author of both papers and developer of the tool (Abstract and author line). Checked against the submission version on 2026-10-06.}, the tool this
paper relies on and, in part, validates. The same person who built the
assessment tool is therefore measuring and reporting on its output.
That includes the validation against external benchmarks
(Section~\ref{sec:validation-power}) and the audit of venue identities
(Section~\ref{sec:identity-misattribution}), which test the tool's
recognition and matching behavior.

We limit this conflict in three ways. First, we did not adjust the
tool's scoring logic based on this paper's results. That covers the
confidence weights, the vote thresholds and the choice of which backends
vote. Correcting the name-based venue-matching assumption is different. An actively
developed research tool ordinarily gets such fixes, and such
corrections do not by themselves compromise this study. Second, every methodological choice made specifically
for this study comes with its rationale: the denominator definitions,
the split into strict and broad, and the leave-one-backend-out analysis.
Third, the tool is open source and the upstream data sources are public
(Section~\ref{sec:data-availability}). Individual classifications can
therefore be re-run with the open-source tool on current public data
instead of being taken on trust. Results for an individual citation may
differ from ours (Section~\ref{sec:reproducibility}). 

\section*{Acknowledgments}
This work builds directly on Aletheia-Probe~\citep{aletheia_probe} and its
underlying data sources. We thank the maintainers of OpenAlex, Crossref,
OpenCitations, DOAJ, DBLP, CORE, PubMed, Scopus, KSCIEN, UGC-CARE,
Beall's List, Retraction Watch, and predatoryjournals.org for making the
underlying data openly available. The author thanks Arndt Kadelka, Matthias
Britsch and Burkhard Alfert for their review and their constructive comments.

\section*{Statements and Declarations}

\textbf{Funding.} This work was supported by the Federal Ministry of
Research, Technology and Space (BMFTR) in Germany under grant number
16KIS2240 of the SUSTAINET-guardian project, of which Deutsche Telekom AG
is a partner.

\textbf{Competing interests.} The author is the developer of
Aletheia-Probe, the tool this paper both relies on and, in part,
validates (Section~\ref{sec:coi}). The author is employed by Deutsche Telekom AG, a partner of the funded project.

\textbf{Data availability.} arXiv's source packages, the bibliometric
databases and predatory lists queried, and the assessment tool itself are all
public; the pipeline's per-citation output is not released, for the reasons given in
Section~\ref{sec:data-availability}.

\textbf{Code availability.} Aletheia-Probe, the assessment tool, is open
source~\citepe{aletheia_probe,florath2026aletheia}{Abstract: open-source tool, repository https://github.com/sustainet-guardian/aletheia-probe. Checked against the submission version on 2026-10-06.}. The scripts that compute this paper's quantitative results are released
separately~\citep{predatory_citations_arxiv_2025_repo}, as code only. Their input, the per-citation output, is not released
(Section~\ref{sec:data-availability}).

\textbf{Author contributions.} The author confirms sole responsibility
for study conception, design, analysis, and manuscript preparation.

\textbf{Ethics approval and consent to participate.} Not applicable:
this study analyzes only publicly available bibliographic metadata and reports aggregate results. It involves no experiments, surveys or interviews with people, and it publishes no results about individual authors.

\appendix

\section{Identity-misattribution forensics}
\label{sec:appendix-identity-forensics}

\subsection{Matching mechanism and identity test}

The nine predatory lists are queried by matching a cited venue's
name against the tool's internal venue records, and four of them also match
by ISSN. The tool builds these records by venue name across the whole corpus
and completes them while it runs. If a citation with an ISSN matches by name a
record that has none, the ISSN is added to the record. Later matches by ISSN
then lead to this record, whether or not it is the same venue.

For each strict-predatory cited reference that carries its own ISSN
(\AuditpopNWithCitingIssn{} of the \AuditpopNOriginalStrict{} original ones),
the identity test (Section~\ref{sec:identity-misattribution}) compares the
ISSN of the citing \texttt{.bib} entry with the ISSN stored for the record
that produced the vote. A mismatch is counted if the stored ISSN is present
and differs, or if it is absent while the citing ISSN resolves cleanly to a
separate record of the same name. Beall's List, the Algerian national list and the KSCIEN predatory lists never extract an ISSN themselves. They win the vote in
\AuditpopNoExtractNWithIssn{} of these cases and account for
\AuditpopNoExtractNMismatch{} of the \AuditpopNMismatch{} mismatches. The two
cloned-journal lists cannot be checked this way, because such a list cannot
tell a genuine journal from an impersonator that uses its ISSN. The
\AuditStructurallyUndecidable{} cited references whose winning vote comes from
one of these lists are therefore structurally undecidable. The test says nothing about whether a predatory list's criteria are
right. It only asks whether the vote's target is the cited venue.

\subsection{Corrected strict-predatory count}
\label{sec:appendix-corrected-count}

Of the \AuditMismatch{} mismatches, \DerivedAuditMismatchMinusSurvivors{} lose
the predatory verdict once the invalid vote is removed and end up as
\texttt{legitimate}, \texttt{suspicious} or \texttt{unknown}. The other
\StrictNSecondVoteSurvivors{} keep it through a second vote from a
cloned-journal list, which is not independent confirmation. Together with the
structurally undecidable references, \StrictUndecided{} cited references in the
corrected count rest solely on such a vote. This paper does not resolve them
either way.

\subsection{Identity-unverified remainder}
\label{sec:appendix-residual-count}

The \DerivedAuditMismatchMinusSurvivors{} mismatch cases plus the
\StrictIdentityUnverified{} identity-unverified matches add up to
\NUnverifiedOfOriginal{} cited references (\PctUnverifiedOfOriginal\% of the
original \AuditOriginalStrictTotal{}) that rest on a weak identity match. This
measures how much of the original count rests on that match, not how much true
prevalence is undercounted (Section~\ref{sec:validation-power}).

\subsection{Case study: Journal of Scientific Computing}

A total of \JscNFlaggedSolelyAlgerian{} cited references to this journal (Springer
Nature, ISSN 0885-7474) match only the Algerian national
list~\citep{DGRSDT2024}, and at least seven other backends outvote that match
with legitimate assessments. The list entry gives the site
\texttt{jscglobal.\allowbreak org}. An Internet Archive snapshot of
2024-02-21 shows that site as \emph{Journal of Scientific Computing} under the
ISSN 1524-2560, which we could not find in Crossref or OpenAlex. We take the
site to be a cloned copy that impersonates the journal. The ISSN or DOI of the
citing entries resolves all \JscNFlaggedSolelyAlgerian{} references to the
genuine journal: \JscrecountGenuineIssnMatchPct\% by ISSN,
\JscrecountGenuineSpringerDoiPct\% by Springer DOI and
\JscrecountGenuineKluwerLegacyDoiPct\% by a pre-2004 Kluwer-legacy DOI. None
carries the clone's ISSN.

\section{Sensitivity analyses}
\label{sec:appendix-sensitivity}

\subsection{Unknown venues: many are well-known names}
Is it fair to assume that \texttt{unknown} venues are no more likely to
be predatory than resolved ones? We can inspect part of that directly. We
ranked the \UnknownvenuesNDistinct{} distinct venues with at least one
\texttt{unknown} citation by how often they are cited. The top 10 alone
account for \UnknownvenuesToptenPct\% of all \UnknownvenuesNTotal{}
\texttt{unknown} cited references, and the top 100 account for
\UnknownvenuesTophundredPct\%. These top venues include many
well-known, clearly legitimate names written in an abbreviated or
non-standard form, such as \emph{APJ}, \emph{MNRAS},
\emph{Phys.\ Rev.\ D/Lett./B/A/C/E/X}, \emph{AAAI}, \emph{NeurIPS},
\emph{ICML}, \emph{ICLR} and \emph{Duke Mathematical Journal}. They are not
obscure or predatory-adjacent.

This check therefore speaks for a little more than half of the
\texttt{unknown} cited references. We did not check the long tail, that is, the
remaining venues and the references they account for. For the top 100, we did not separate venues that no backend indexes from venues that were not matched because of the name form. 

Fixing this by hand is not realistic. In our own curated list of name
variants~\citepe{venue_acronyms_2025}{Public dataset file dataset/venue-acronyms-2025-curated.json (version 2.0, generated 2026-02, 6,540 acronyms): the AAAI entry lists 177 name variants, including one unrelated conference (CVPR). Counted by the author's list and by script on 2026-10-07.}, collected during this evaluation,
AAAI alone has more than 170 variants. Building such a list takes real
judgment calls, for example whether a co-located workshop is the same
\emph{venue} as its parent conference, and even this list has its own
errors: one entry names an unrelated conference. So the pattern above
should be read narrowly. That does not mean that resolving these venues, or
any match already resolved, is guaranteed correct.

\subsection{Validating the replay}
Before trusting our leave-one-backend-out re-implementation, we checked
that it reproduces the tool's own stored verdict exactly when no
backend is excluded. It achieves this: \LeaveOneOutReimplementationMatchPct{}\% match, on all \BroadCount{}
broad-predatory entries.

We validate against these \BroadCount{} entries only. The full
\CorpusConfirmedCitedEntries{} confirmed-cited entries are not needed.
Aletheia-Probe's decision rule (Section~\ref{sec:methodology}) never
reaches a \texttt{predatory} verdict unless predatory-list evidence wins
the confidence-weighted comparison of Section~\ref{sec:assessment-logic}, and this holds whichever single
backend is excluded, even a legitimate-side one like Scopus. That is
exactly how the \BroadCount{}-entry set is defined. No entry outside it
can therefore become \texttt{predatory} by excluding one backend, and
there is nothing left to check in the remaining entries.

\subsection{Which backends matter most}
Removing the Algerian national list alone has the biggest effect of any
single backend on the broad-predatory count (for the strict count, excluding
Scopus moves it more, Section~\ref{sec:leave-one-out}). The broad-predatory count drops from \BroadCount{} to
\LeaveOneOutAlgerianMinistryBroadAfterExclusion{}
(\LeaveOneOutAlgerianMinistryBroadPctChange\%), because every citation
whose \emph{only} predatory-list hit came from this source falls out
entirely. The strict-predatory count drops from \StrictCount{} to
\LeaveOneOutAlgerianMinistryStrictAfterExclusion{} under the same test,
so \DerivedAlgerianExclusionStrictLossN{} cited references lose their \texttt{predatory} verdict. The same
weighing is behind the \emph{Journal of Scientific Computing} case
(Appendix~\ref{sec:appendix-identity-forensics}): a single
high-confidence predatory-list match is set against several lower-confidence
legitimate matches for one venue, and across the corpus these cases add
up.

Every other predatory list moves the broad-predatory count far less when
removed on its own. For KSCIEN's standalone-journals list the change is
\LeaveOneOutKscienStandaloneJournalsBroadPctChange\%, and for all others
\LeaveOneOutUgcCareClonedGrouptwoBroadPctChange\% or less.
So the Algerian national list is not only the list with the lowest
second-vote share (Table~\ref{tab:reliability}). It is also, by a wide
margin, the one whose removal moves the results most.

Apart from Scopus (Section~\ref{sec:leave-one-out}), the analyzers and the other legitimate lists matter less individually.
In a separate replay over the \LooexhaustivePopulationSize{} confirmed-cited
references with conflicting evidence (any verdict change counted), removing OpenAlex flips
\LooexhaustiveOpenalexAnalyzerFlipPct\% of the cases considered, while
removing Crossref or OpenCitations flips
\LooexhaustiveCrossrefAnalyzerFlipPct\% and
\LooexhaustiveOpencitationsAnalyzerFlipPct\%. DBLP is the one exception.
It returns a match for only \LooexhaustiveDblpVenuesNConsidered{} of the
\LooexhaustivePopulationSize{} conflicted entries, but
\LooexhaustiveDblpVenuesFlipPct\% of those flip. It matches rarely, but when it
matches it often changes the verdict.

\subsection{Mechanisms behind backend-removal effects}
 KSCIEN's list of hijacked
journals and Retraction Watch are structurally excluded from the
confidence-weighted vote (Section~\ref{sec:methodology}), so removing
either changes nothing (\StrictCount{}, unchanged). Neither is ever the
invalidated backend in any of the \AuditMismatch{} audit cases.

Our leave-one-backend-out replay excludes not only the removed backend but also
any backend that our identity-misattribution audit
(Section~\ref{sec:identity-misattribution}) already found invalid for
that citation. That applies to all
\AuditMismatch{} cited references with a mismatching vote, including the
\StrictNSecondVoteSurvivors{} of Appendix~\ref{sec:appendix-corrected-count}. Without this, an
already-invalidated vote could win by default, including the survivors' own invalidated first vote.

For these \StrictNSecondVoteSurvivors{}, only the UGC-CARE cloned list (Group II) matters.
It supplies the \StrictNSecondVoteSurvivors{} survivors'
only remaining valid vote, so removing it strips their last piece of
predatory-list evidence too:
\LeaveOneOutUgcCareClonedGrouptwoStrictAfterExclusion{}. That is still
inside the established leave-one-backend-out range
(\LeaveOneOutAlgerianMinistryStrictAfterExclusion{}--\LeaveOneOutScopusStrictAfterExclusion{})
and not a new extreme. Section~\ref{sec:limitations} describes a related
effect of the same clone pages.

In total, \LoNFlippedTotal{} cited references flip, not just these \StrictNSecondVoteSurvivors{}. The other
\LoNIndependent{} have nothing to do with the second-vote-survivor
mechanism. Of them, \LoNSoleSurvivingVote{} rely on
the UGC-CARE cloned list (Group II) as
their only predatory-list evidence, and the remaining
\LoNRazorThinTie{} are near-ties against a genuinely valid
vote of the Algerian national list, where removing either vote
alone flips the verdict to \texttt{legitimate}.

Removing Scopus works the other way. Scopus currently supplies
legitimate-side weight that tips some entries away from
\texttt{predatory}. Without that counterweight, predatory-list evidence
wins instead, which is why excluding Scopus \emph{raises} the
strict-predatory count (Table~\ref{tab:strict-alternatives}).

\section{Disciplinary composition}
\label{sec:appendix-disciplinary}

\subsection{How we stratified}
Section~\ref{sec:disciplinary} raises a specific hypothesis: arXiv's own
disciplinary mix, dominated by computer science, physics and mathematics
rather than some of the disciplines more strongly represented in prior work, explains the low rate we observe. We examined this instead of
asserting it. We split the per-submission predatory-citation
rates (\SubmissionStrictPct\% strict, \SubmissionBroadPct\% broad;
Section~\ref{sec:headline-prevalence}) by each submission's primary
arXiv subject category. We group
\texttt{eess}, \texttt{stat}, \texttt{q-bio}, \texttt{econ},
\texttt{q-fin}, \texttt{math-ph} and \texttt{nlin} together as
``Other''. All \DiscCorpuswideNSubmissions{} submissions resolve to a
category from arXiv's own Open Archives Initiative (OAI) metadata snapshot (100\% resolved).
Section~\ref{sec:disciplinary-breakdown} reports the result
(Table~\ref{tab:disciplinary}).

\subsection{The Mathematics anomaly}
Mathematics is the one group whose broad rate is well above the corpus-wide
baseline. Per submission, the rate is \DiscMathBroadPct\% against
\SubmissionBroadPct\% (\DerivedMathSubmissionBroadMultiple{}$\times$). Per cited reference, it is
\DiscciteMathBroadPct\% against \MultibibCurrentBroadRate\%
(\MathBroadMultipleCorrected$\times$).

But this is not a broad disciplinary pattern. \MathjscPctJsc\% of
Mathematics's broad-predatory cited references trace to one venue, \emph{Journal of Scientific Computing}, which the
tool's full-evidence assessment classifies as \texttt{legitimate} for every
citation. Exclude this one venue, and Mathematics's citation-level broad rate
falls to \MathjscRateExclJscPct\%, below the corpus-wide baseline.

Mathematics is also underrepresented here. It makes up only
\DisccompinclMathPct\% of included submissions, against
\DisccompexclMathPct\% of the submissions that this study's BibTeX-based
coverage excludes (Section~\ref{sec:corpus}). The result therefore
describes the minority of arXiv Mathematics output that uses a separate
\texttt{.bib} file, not Mathematics submissions in general.

\section{Benchmark validation}
\label{sec:appendix-benchmark}

Section~\ref{sec:add-one} simulates adding two predatory-side lists as new
backends. This appendix asks the reverse: does the tool already recognize
the venues on them? Existing predatory lists already cover \AddoneRwCoveragePct\% of
the Hijacked Journal Checker cases and \AddoneDoajCoveragePct\% of the DOAJ
withdrawals. For the rest, a non-predatory verdict follows from the definition,
since no predatory list matches them. What matters is how the tool labels them
instead.

For DOAJ, \DoajxtabLegitPctOfDisjoint{}\% of the withdrawn journals that no predatory list matches are still called
\texttt{legitimate}. Of those, \DoajxtabLegitPctReinstated\% have since been
reinstated by DOAJ, about double the \DoajxtabNonlegitPctReinstated\% among the
others. Across all \DoajReinclusionNCheckable{} withdrawn journals that could
be checked, the share is \DoajReinclusionPctReinstated\%
(Section~\ref{sec:add-one}). DOAJ withdraws journals for missing best-practice standards, which is
broader than being predatory. The figure therefore stress-tests the
\texttt{legitimate} label and is not an estimate of false negatives.

For the Hijacked Journal Checker the result cannot be interpreted. A hijacked journal's
cloned site often uses the real journal's ISSN (\RwcloneExactStringMatchN{} of
\RwcloneNTotal{} cases exactly, \RwcloneOverlappingNormalizedMatchN{} more in an overlapping
form; \RwcloneHijackedIssnMissingN{} have no clone ISSN recorded), so a \texttt{legitimate} verdict may be right or wrong.
Only six cases have a different ISSN for the clone, too few to test.

\section{Corpus and population characterization}
\label{sec:appendix-population}

\subsection{Venue type, entry type and DOI presence}
Predatory matching is heavily concentrated in journal-type venues.
\VenuetypeJournalBroadPct\% of the \DerivedVenuetypeJournalMillions{} million
\texttt{journal}-type cited references are broad-predatory. For the \DerivedVenuetypeConferenceMillions{} million cited references to
\texttt{conference}-type venues the share is
\VenuetypeConferenceBroadPct\%, roughly 20 times lower. For workshop or
proceedings cited references it is 0\%.

This gap may be inflated, though, by an asymmetry in our own ground
truth, so it does not purely reflect prevalence. Of the nine predatory lists
in Table~\ref{tab:reliability}, only the KSCIEN predatory-conferences
list targets conferences specifically. It is among the smallest of the nine by match count and
among the least corroborated (Table~\ref{tab:reliability}), and it
contributes well under 1\% of all predatory-list match instances. The true
predatory-conference rate is therefore confounded with our detection
coverage and cannot be cleanly separated from it using this tool.

By BibTeX entry type the same pattern holds. The broad-predatory rate
is \EntrytypeInproceedingsBroadPct\% for conference papers (BibTeX type
\emph{inproceedings}) and
\EntrytypeArticleBroadPct\% for journal articles (type
\emph{article}). For every other entry type
with a non-trivial count it is 0\%.

DOI presence points in the direction one would expect, with a much
smaller gap. Legitimate cited references carry a DOI
\DoiLegitimatePct\% of the time, against \DoiStrictPredatoryPct\% for
strict-predatory ones (a \DerivedDoiGapPoints{}-point gap). The gap is narrow enough that
DOI presence alone should not be read as a strong discriminator between
the two groups.

\subsection{Citation intensity}
Using the source-extracted citation-log join
(Section~\ref{sec:citation-log}), we count how many times each
confirmed-cited entry's key appears in a
\texttt{\textbackslash cite\{\}}-style command across the paper's
\LaTeX{} source. At least half of the references in every assessment category are cited exactly once, so the median is 1. The mean is modestly higher for
\texttt{legitimate} (\IntensityLegitimateMean{}) than for
\texttt{predatory} cited references (\IntensityPredatoryMean{}). That is consistent with
predatory cited references skewing toward single, passing mentions, but
we treat it as a first-pass measurement and not as evidence for any
particular explanation.

\clearpage

  \balance
  \onecolumn
  \printbibliography

\end{document}

%% file: numbers.tex
\newcommand{\AddoneAclDistinctVenues}{2{,}918}
\newcommand{\AddoneAclMovedSuspiciousToLegitimateN}{11}
\newcommand{\AddoneAclMovedToLegitimateN}{66{,}144}
\newcommand{\AddoneAclNAffected}{90{,}947}
\newcommand{\AddoneAclNConflicts}{0}
\newcommand{\AddoneAclPctUnknownSuspicious}{72.7}
\newcommand{\AddoneAclStrictDelta}{0}
\newcommand{\AddoneCombinedBroadDelta}{11{,}165}
\newcommand{\AddoneCombinedNAffected}{102{,}643}
\newcommand{\AddoneCombinedNMultiSource}{89}
\newcommand{\AddoneCombinedStrictDelta}{4{,}281}
\newcommand{\AddoneDoajBroadDelta}{9{,}752}
\newcommand{\AddoneDoajCoveragePct}{4.9}
\newcommand{\AddoneDoajNAffected}{9{,}995}
\newcommand{\AddoneDoajNConflicts}{2{,}543}
\newcommand{\AddoneDoajNPredatoryKept}{118}
\newcommand{\AddoneDoajStrictDelta}{3{,}859}
\newcommand{\AddoneRwBroadDelta}{1{,}442}
\newcommand{\AddoneRwCoveragePct}{53.4}
\newcommand{\AddoneRwNAffected}{1{,}790}
\newcommand{\AddoneRwNConflicts}{221}
\newcommand{\AddoneRwNPredatoryKept}{15}
\newcommand{\AddoneRwStrictDelta}{358}
\newcommand{\AlgerianBroadSharePct}{69.8}
\newcommand{\AuditCheckableWithIssn}{1{,}357}

\newcommand{\AuditMismatch}{534}
\newcommand{\AuditMismatchRateOfCheckable}{39.4}
\newcommand{\AuditNoIssn}{1{,}366}
\newcommand{\AuditNoIssnPctOfRaw}{50.2}
\newcommand{\AuditOriginalStrictTotal}{2{,}723}
\newcommand{\AuditStructurallyUndecidable}{6}

\newcommand{\AuditpopNMismatch}{534}
\newcommand{\AuditpopNOriginalStrict}{2{,}723}

\newcommand{\AuditpopNWithCitingIssn}{1{,}357}

\newcommand{\AuditpopNoExtractNMismatch}{510}

\newcommand{\AuditpopNoExtractNWithIssn}{1{,}294}

\newcommand{\BackendsSplitPctExactlyOne}{60.2}

\newcommand{\BblaccuracyPrecisionPctShort}{97.1}

\newcommand{\BblaccuracyRecallPctShort}{97.5}
\newcommand{\BroadCount}{5{,}911}

\newcommand{\BulkAclFlaggedVenues}{19}

\newcommand{\BulkCryptobibDistinctVenues}{48}

\newcommand{\BulkCryptobibFlaggedVenues}{3}

\newcommand{\ConcLegitDistinctVenues}{33{,}654}
\newcommand{\ConcLegitEffectiveVenues}{120}
\newcommand{\ConcLegitGini}{0.94}
\newcommand{\ConcLegitHhi}{0.008}
\newcommand{\ConcLegitTopfiftyPct}{39.9}
\newcommand{\ConcLegitToptenPct}{21.9}
\newcommand{\ConcLegitTotalCitations}{4{,}175{,}927}
\newcommand{\ConcLegitVenuesForFiftypctN}{110}
\newcommand{\ConcLegitVenuesForFiftypctPct}{0.3}
\newcommand{\ConcLegitVenuesForNinetypctN}{2{,}524}
\newcommand{\ConcLegitVenuesForNinetypctPct}{7.5}
\newcommand{\ConcStrictDistinctVenues}{343}
\newcommand{\ConcStrictEffectiveVenues}{26}
\newcommand{\ConcStrictGini}{0.70}
\newcommand{\ConcStrictHhi}{0.038}
\newcommand{\ConcStrictTopfiftyPct}{68.5}
\newcommand{\ConcStrictToptenPct}{40.5}
\newcommand{\ConcStrictVenuesForFiftypctN}{20}
\newcommand{\ConcStrictVenuesForFiftypctPct}{5.8}
\newcommand{\ConcStrictVenuesForNinetypctN}{161}
\newcommand{\ConcStrictVenuesForNinetypctPct}{46.9}
\newcommand{\ConcernBroadOnlyPct}{27.5}
\newcommand{\ConcernBroadTotalPct}{33.9}
\newcommand{\ConcernHijackedOnlyPct}{2.4}
\newcommand{\ConcernHijackedTotalPct}{3.2}
\newcommand{\ConcernNAnyFlag}{13{,}847}

\newcommand{\ConcernNPapers}{166{,}761}

\newcommand{\ConcernPctRetractionOfFlagged}{69.7}
\newcommand{\ConcernRetractionOnlyPct}{63.3}

\newcommand{\ConfpopNMissedKeys}{221{,}295}

\newcommand{\ConfpopNOwnLog}{8{,}481{,}558}

\newcommand{\ConfpopNPromotedNoLog}{17{,}919}
\newcommand{\ConfpopNPromotedNoLogIsolated}{15{,}720}
\newcommand{\ConfpopNPromotedNoLogSibling}{2{,}199}

\newcommand{\CorpusAssessedFiles}{189{,}637}
\newcommand{\CorpusCandidatePopulation}{27{,}583{,}307}
\newcommand{\CorpusConfirmedCitedEntries}{8{,}720{,}772}
\newcommand{\CorpusCoveredSubmissions}{167{,}727}
\newcommand{\CorpusCoveredSubmissionsPct}{58.96}

\newcommand{\CorpusEmptyFiles}{283}
\newcommand{\CorpusExcludedEntries}{18{,}570{,}868}
\newcommand{\CorpusExcludedSubmissions}{116{,}759}
\newcommand{\CorpusExcludedSubmissionsPct}{41.0}
\newcommand{\CorpusMacroonlyFiles}{3{,}577}
\newcommand{\CorpusMaxEntriesPerSubmission}{113{,}114}
\newcommand{\CorpusMegaRawEntries}{95{,}091{,}424}
\newcommand{\CorpusMegaRawEntriesPct}{83.96}
\newcommand{\CorpusMultiBibExcessDupRows}{164{,}130}

\newcommand{\CorpusMultiBibFilesPct}{19.9}
\newcommand{\CorpusMultiBibPct}{9.45}

\newcommand{\CorpusNMultiBibSubmissions}{15{,}853}
\newcommand{\CorpusNonbibliographyFiles}{3{,}860}
\newcommand{\CorpusNosidecarFiles}{10{,}975}
\newcommand{\CorpusNosidecarIsolatedFilesN}{1{,}381}
\newcommand{\CorpusNosidecarIsolatedPct}{12.6}
\newcommand{\CorpusNosidecarSiblingFilesN}{9{,}594}
\newcommand{\CorpusNosidecarSiblingPct}{87.4}
\newcommand{\CorpusRawTotalEntries}{113{,}262{,}987}
\newcommand{\CorpusSidecarFiles}{178{,}662}
\newcommand{\CorpusTotalSubmissions}{284{,}486}
\newcommand{\CorpusTotalBibFiles}{193{,}497}
\newcommand{\CorpusUnverifiedEntries}{291{,}667}

\newcommand{\CorrectionBeforeBroadCitationsN}{6{,}430}

\newcommand{\CoverageUnknownOfEligiblePct}{34.7}
\newcommand{\DerivedAddoneCombinedBroadTotal}{17{,}076}
\newcommand{\DerivedAddoneCombinedStrictTotal}{6{,}478}
\newcommand{\DerivedAddoneDoajBroadTotal}{15{,}663}
\newcommand{\DerivedAddoneDoajStrictTotal}{6{,}056}
\newcommand{\DerivedAddoneRwBroadTotal}{7{,}353}
\newcommand{\DerivedAddoneRwStrictTotal}{2{,}555}
\newcommand{\DerivedAlgerianExclusionStrictLossN}{363}
\newcommand{\DerivedAlgerianExclusionStrictMultiple}{0.83}
\newcommand{\DerivedAlgerianMinistryNotRetainedPct}{80.2}
\newcommand{\DerivedAssessedPctDisplay}{75.1}
\newcommand{\DerivedAssessedPctOfBibFiles}{98.01}
\newcommand{\DerivedAuditMismatchMinusSurvivors}{526}
\newcommand{\DerivedAuditMismatchPctOfOriginal}{19.6}
\newcommand{\DerivedBloatedPctOfPairwiseTotal}{2.0}
\newcommand{\DerivedConcernPctOfPapers}{8.30}
\newcommand{\DerivedConfirmedCitedPctOfCandidate}{31.62}
\newcommand{\DerivedCorrectionSizePct}{19.3}

\newcommand{\DerivedDoiGapPoints}{7.7}
\newcommand{\DerivedEmptyPctOfBibFiles}{0.15}
\newcommand{\DerivedExcludedPctOfCandidate}{68.4}
\newcommand{\DerivedExcludedPctOfCandidateOnly}{67.33}
\newcommand{\DerivedExcludedPlusUnverifiedN}{18{,}862{,}535}
\newcommand{\DerivedHeadlineRowsSumPct}{100.0}
\newcommand{\DerivedLegitHhiDisplayedReciprocal}{125}
\newcommand{\DerivedLegitPlusUnknownDistinctVenues}{231{,}902}
\newcommand{\DerivedLegitPlusUnknownStrictMultiple}{2{,}934}
\newcommand{\DerivedLegitStrictMultiple}{1{,}901}
\newcommand{\DerivedLegitUnknownVenuesPerStrictVenue}{676}

\newcommand{\DerivedMacroonlyPctOfBibFiles}{1.85}
\newcommand{\DerivedMathSubmissionBroadMultiple}{2.0}
\newcommand{\DerivedNoVenuePctDisplay}{1.5}
\newcommand{\DerivedNonbibliographyPctOfBibFiles}{1.99}
\newcommand{\DerivedNosidecarPctDisplay}{5.79}
\newcommand{\DerivedPrecisionSampleAtPrintedRateN}{49{,}000}
\newcommand{\DerivedPreprintPctDisplay}{15.3}
\newcommand{\DerivedPrimaryStrictRatePrinted}{0.00051}
\newcommand{\DerivedRuleOfThreeAtPrintedRateN}{5{,}880}
\newcommand{\DerivedScopusExclusionStrictMultiple}{1.22}
\newcommand{\DerivedSidecarPctOfAssessedFiles}{94.21}
\newcommand{\DerivedSisfIssnlStrictMaxMovementPct}{0.8}
\newcommand{\DerivedStrictvoteGeResolvedCorrectedMultiple}{6.10}
\newcommand{\DerivedStrictvoteVotesGeLegitResolvedCorrectedMultiple}{71.36}
\newcommand{\DerivedSubmissionsWithoutConfirmedCitedN}{966}
\newcommand{\DerivedUnknownStrictMultiple}{1{,}033}
\newcommand{\DerivedUnsupportedTypePctDisplay}{8.1}
\newcommand{\DerivedUnverifiedPctOfCandidate}{1.06}
\newcommand{\DerivedValidationFnHundredthPctMultiple}{2.9}
\newcommand{\DerivedValidationFnHundredthPctN}{645}
\newcommand{\DerivedValidationFnOnePctMultiple}{293.0}
\newcommand{\DerivedValidationFnOnePctN}{64{,}456}
\newcommand{\DerivedValidationFnTenthPctMultiple}{29.3}
\newcommand{\DerivedValidationFnTenthPctN}{6{,}446}
\newcommand{\DerivedValidationFnThousandthPctMultiple}{0.3}
\newcommand{\DerivedValidationFnThousandthPctN}{64}
\newcommand{\DerivedValidationFpTenPctMultiple}{1.0}
\newcommand{\DerivedValidationFpTenPctN}{220}
\newcommand{\DerivedVenuetypeConferenceMillions}{1.72}

\newcommand{\DerivedVenuetypeJournalMillions}{4.64}
\newcommand{\DiscCorpuswideBroadCiHi}{2.90}
\newcommand{\DiscCorpuswideBroadCiLo}{2.74}
\newcommand{\DiscCorpuswideBroadPct}{2.82}
\newcommand{\DiscCorpuswideMeanEntries}{52.3}
\newcommand{\DiscCorpuswideNSubmissions}{166{,}761}
\newcommand{\DiscCorpuswidePctOfCorpus}{100.0}
\newcommand{\DiscCorpuswideStrictCiHi}{1.20}
\newcommand{\DiscCorpuswideStrictCiLo}{1.10}
\newcommand{\DiscCorpuswideStrictPct}{1.15}
\newcommand{\DiscCsBroadCiHi}{3.13}
\newcommand{\DiscCsBroadCiLo}{2.92}
\newcommand{\DiscCsBroadPct}{3.02}
\newcommand{\DiscCsMeanEntries}{49.6}
\newcommand{\DiscCsNSubmissions}{96{,}771}
\newcommand{\DiscCsPctOfCorpus}{58.0}
\newcommand{\DiscCsStrictCiHi}{1.50}
\newcommand{\DiscCsStrictCiLo}{1.35}

\newcommand{\DiscCsStrictPct}{1.42}
\newcommand{\DiscMathBroadCiHi}{6.07}
\newcommand{\DiscMathBroadCiLo}{5.24}
\newcommand{\DiscMathBroadPct}{5.64}
\newcommand{\DiscMathMeanEntries}{34.9}
\newcommand{\DiscMathNSubmissions}{12{,}023}
\newcommand{\DiscMathPctOfCorpus}{7.2}
\newcommand{\DiscMathStrictCiHi}{0.51}
\newcommand{\DiscMathStrictCiLo}{0.29}

\newcommand{\DiscMathStrictPct}{0.38}
\newcommand{\DiscOtherBroadCiHi}{3.06}
\newcommand{\DiscOtherBroadCiLo}{2.56}
\newcommand{\DiscOtherBroadPct}{2.80}
\newcommand{\DiscOtherMeanEntries}{42.4}
\newcommand{\DiscOtherNSubmissions}{16{,}812}
\newcommand{\DiscOtherPctOfCorpus}{10.1}
\newcommand{\DiscOtherStrictCiHi}{1.36}
\newcommand{\DiscOtherStrictCiLo}{1.03}

\newcommand{\DiscOtherStrictPct}{1.18}
\newcommand{\DiscPhysicsBroadCiHi}{1.64}
\newcommand{\DiscPhysicsBroadCiLo}{1.40}
\newcommand{\DiscPhysicsBroadPct}{1.51}
\newcommand{\DiscPhysicsMeanEntries}{67.8}
\newcommand{\DiscPhysicsNSubmissions}{41{,}155}
\newcommand{\DiscPhysicsPctOfCorpus}{24.7}
\newcommand{\DiscPhysicsStrictCiHi}{0.81}
\newcommand{\DiscPhysicsStrictCiLo}{0.64}

\newcommand{\DiscPhysicsStrictPct}{0.72}
\newcommand{\DiscciteCsStrictPct}{0.033}
\newcommand{\DiscciteMathBroadPct}{0.287}
\newcommand{\DiscciteMathStrictPct}{0.012}
\newcommand{\DiscciteOtherStrictPct}{0.032}
\newcommand{\DisccitePhysicsStrictPct}{0.012}
\newcommand{\DisccompexclCsPct}{27.2}
\newcommand{\DisccompexclMathPct}{26.2}
\newcommand{\DisccompexclPhysicsPct}{36.6}
\newcommand{\DisccompinclCsPct}{57.9}
\newcommand{\DisccompinclMathPct}{7.2}
\newcommand{\DisccompinclPhysicsPct}{24.8}
\newcommand{\DiscsubcountNBroad}{4{,}697}
\newcommand{\DiscsubcountNStrict}{1{,}917}
\newcommand{\DiscsubcountNTotal}{166{,}761}
\newcommand{\DoajReinclusionLegitresolvingPctStillRemoved}{79.6}
\newcommand{\DoajReinclusionNCheckable}{4{,}264}
\newcommand{\DoajReinclusionPctReinstated}{16.25}

\newcommand{\DoajxtabLegitPctOfDisjoint}{67.4}
\newcommand{\DoajxtabLegitPctReinstated}{20.4}

\newcommand{\DoajxtabNonlegitPctReinstated}{9.6}

\newcommand{\DoiLegitimatePct}{30.9}
\newcommand{\DoiStrictPredatoryPct}{23.2}
\newcommand{\EligiblePopulation}{6{,}547{,}551}
\newcommand{\EntrytypeArticleBroadPct}{0.100}
\newcommand{\EntrytypeInproceedingsBroadPct}{0.007}
\newcommand{\HeadlineHijackedN}{573}
\newcommand{\HeadlineHijackedPct}{0.007}
\newcommand{\HeadlineInsufficientDataN}{1{,}236}
\newcommand{\HeadlineInsufficientDataPct}{0.014}
\newcommand{\HeadlineLegitimateN}{4{,}175{,}927}
\newcommand{\HeadlineLegitimatePct}{47.9}
\newcommand{\HeadlineNotApplicableN}{2{,}173{,}221}
\newcommand{\HeadlineNotApplicablePct}{24.9}
\newcommand{\HeadlineSuspiciousN}{98{,}518}
\newcommand{\HeadlineSuspiciousPct}{1.1}
\newcommand{\HeadlineTotal}{8{,}720{,}772}
\newcommand{\HeadlineUnknownN}{2{,}269{,}673}
\newcommand{\HeadlineUnknownPct}{26.0}
\newcommand{\HijackedbenchDoajTotal}{4{,}352}
\newcommand{\HijackedbenchRwTotal}{455}
\newcommand{\IntensityLegitimateMean}{1.95}
\newcommand{\IntensityPredatoryMean}{1.46}
\newcommand{\IssnpresenceGapPostcorr}{21.0}
\newcommand{\IssnpresenceLegitPct}{58.8}
\newcommand{\IssnpresencePostcorrPct}{37.8}
\newcommand{\IssnpresencePostcorrWithIssnN}{831}
\newcommand{\JscNFlaggedSolelyAlgerian}{1{,}719}

\newcommand{\JscrecountGenuineIssnMatchPct}{73.2}
\newcommand{\JscrecountGenuineKluwerLegacyDoiPct}{0.7}
\newcommand{\JscrecountGenuineSpringerDoiPct}{26.1}
\newcommand{\LoNFlippedTotal}{77}
\newcommand{\LoNIndependent}{69}
\newcommand{\LoNRazorThinTie}{66}
\newcommand{\LoNSoleSurvivingVote}{3}

\newcommand{\LeaveOneOutAlgerianMinistryBroadAfterExclusion}{2{,}961}
\newcommand{\LeaveOneOutAlgerianMinistryBroadPctChange}{$-$49.9}
\newcommand{\LeaveOneOutAlgerianMinistryStrictAfterExclusion}{1{,}834}
\newcommand{\LeaveOneOutAlgerianMinistryStrictPctChange}{$-$16.5}

\newcommand{\LeaveOneOutKscienStandaloneJournalsBroadPctChange}{$-$5.3}
\newcommand{\LeaveOneOutReimplementationMatchPct}{100.0}

\newcommand{\LeaveOneOutScopusStrictAfterExclusion}{2{,}680}
\newcommand{\LeaveOneOutScopusStrictPctChange}{22.0}

\newcommand{\LeaveOneOutUgcCareClonedGrouptwoBroadPctChange}{$-$2.0}
\newcommand{\LeaveOneOutUgcCareClonedGrouptwoStrictAfterExclusion}{2{,}120}
\newcommand{\LegitreclassifiedN}{457}
\newcommand{\LegitreclassifiedPct}{0.011}
\newcommand{\LooexhaustiveCrossrefAnalyzerFlipPct}{0.1}
\newcommand{\LooexhaustiveDblpVenuesFlipPct}{33.7}
\newcommand{\LooexhaustiveDblpVenuesNConsidered}{32{,}406}
\newcommand{\LooexhaustiveOpenalexAnalyzerFlipPct}{5.6}
\newcommand{\LooexhaustiveOpencitationsAnalyzerFlipPct}{0.4}
\newcommand{\LooexhaustivePopulationSize}{659{,}588}
\newcommand{\MathBroadMultipleCorrected}{4.2}

\newcommand{\MathjscPctJsc}{91.4}
\newcommand{\MathjscRateExclJscPct}{0.025}
\newcommand{\MegafamAclPct}{91.68}
\newcommand{\MegafamCryptobibPct}{2.94}
\newcommand{\MegafamOtherPct}{5.38}
\newcommand{\MirrorexhaustiveNCheckable}{2{,}296}
\newcommand{\MirrorexhaustiveNTrueCandidates}{208{,}569}
\newcommand{\MirrorexhaustivePctConfirmedMisattribution}{55.0}
\newcommand{\MirrorexhaustivePctNoCitingIssn}{98.90}
\newcommand{\MissedkeysCurrentBroadAfterN}{80}

\newcommand{\MissedkeysCurrentStrictAfterN}{51}

\newcommand{\MissedkeysNPopulation}{221{,}295}

\newcommand{\MultibibBroadNExcessRows}{110}
\newcommand{\MultibibBroadPctOfPopulation}{1.86}

\newcommand{\MultibibCurrentBroadRate}{0.068}
\newcommand{\MultibibCurrentStrictRate}{0.025}
\newcommand{\MultibibStrictNExcessRows}{33}
\newcommand{\MultibibStrictPctOfPopulation}{1.50}

\newcommand{\NUnverifiedOfOriginal}{1{,}310}

\newcommand{\NociteBblOnlyAttributable}{12{,}719}
\newcommand{\NociteBblOnlyAttributablePct}{31.5}
\newcommand{\NociteBblOnlyNonAttributable}{27{,}638}
\newcommand{\NociteBblOnlyTotal}{40{,}357}
\newcommand{\NociteExcludedBroadCount}{5{,}903}
\newcommand{\NociteExcludedBroadRate}{0.068}
\newcommand{\NociteExcludedDenominator}{8{,}710{,}316}
\newcommand{\NociteExcludedStrictCount}{2{,}192}
\newcommand{\NociteExcludedStrictRate}{0.025}
\newcommand{\NociteNPromotedAttributable}{10{,}456}
\newcommand{\NociteNPromotedAttributableBroad}{8}
\newcommand{\NociteNPromotedAttributableStrict}{5}
\newcommand{\NociteNPromotedTotal}{239{,}214}
\newcommand{\NocitePctBroadOfAttributable}{0.077}
\newcommand{\NocitePctOfPromotedAttributable}{4.4}
\newcommand{\NocitePctStrictOfAttributable}{0.048}
\newcommand{\NosidecarcorrABroadN}{5{,}911}
\newcommand{\NosidecarcorrABroadPct}{0.068}

\newcommand{\NosidecarcorrAStrictN}{2{,}197}

\newcommand{\NosidecarcorrBBroadPct}{0.014}

\newcommand{\NosidecarcorrBStrictPct}{0.007}

\newcommand{\NosidecarcorrCBroadN}{5{,}951}
\newcommand{\NosidecarcorrCBroadPct}{0.066}

\newcommand{\NosidecarcorrCStrictN}{2{,}218}

\newcommand{\PairwiseCrOcBothFound}{4{,}193{,}470}
\newcommand{\PairwiseCrOcRate}{87.0}
\newcommand{\PairwiseNBloatedFiles}{3{,}843}
\newcommand{\PairwiseNNormalFiles}{185{,}794}
\newcommand{\PairwiseOaCrBothFound}{4{,}903{,}314}
\newcommand{\PairwiseOaCrRate}{81.3}
\newcommand{\PairwiseOaOcBothFound}{4{,}152{,}551}
\newcommand{\PairwiseOaOcRate}{99.4}
\newcommand{\PctUnverifiedOfOriginal}{48.1}
\newcommand{\PrimaryStrictRatePct}{0.051}

\newcommand{\ReevalsampleStrictMainN}{4}

\newcommand{\ReliabilityAlgerianMinistryCoFlagPct}{37.6}
\newcommand{\ReliabilityAlgerianMinistryFlags}{4{,}125}
\newcommand{\ReliabilityAlgerianMinistryLeaveFocalOutPct}{19.8}
\newcommand{\ReliabilityBeallsCoFlagPct}{99.3}
\newcommand{\ReliabilityBeallsFlags}{2{,}320}
\newcommand{\ReliabilityBeallsLeaveFocalOutPct}{63.6}
\newcommand{\ReliabilityKscienPredatoryConferencesCoFlagPct}{94.3}
\newcommand{\ReliabilityKscienPredatoryConferencesFlags}{53}
\newcommand{\ReliabilityKscienPredatoryConferencesLeaveFocalOutPct}{0.0}
\newcommand{\ReliabilityKscienPublishersCoFlagPct}{100.0}
\newcommand{\ReliabilityKscienPublishersFlags}{102}
\newcommand{\ReliabilityKscienPublishersLeaveFocalOutPct}{98.0}
\newcommand{\ReliabilityKscienStandaloneJournalsCoFlagPct}{99.6}
\newcommand{\ReliabilityKscienStandaloneJournalsFlags}{2{,}202}
\newcommand{\ReliabilityKscienStandaloneJournalsLeaveFocalOutPct}{61.9}
\newcommand{\ReliabilityPredatoryjournalsCoFlagPct}{96.7}
\newcommand{\ReliabilityPredatoryjournalsFlags}{121}
\newcommand{\ReliabilityPredatoryjournalsLeaveFocalOutPct}{94.2}
\newcommand{\ReliabilityUgcCareClonedCoFlagPct}{100.0}
\newcommand{\ReliabilityUgcCareClonedFlags}{7}
\newcommand{\ReliabilityUgcCareClonedGroupCoFlagPct}{94.3}
\newcommand{\ReliabilityUgcCareClonedGroupFlags}{229}
\newcommand{\ReliabilityUgcCareClonedGroupLeaveFocalOutPct}{2.2}
\newcommand{\ReliabilityUgcCareClonedLeaveFocalOutPct}{0.0}
\newcommand{\ReliabilityUgcCareDelistedGroupCoFlagPct}{60.0}
\newcommand{\ReliabilityUgcCareDelistedGroupFlags}{30}
\newcommand{\ReliabilityUgcCareDelistedGroupLeaveFocalOutPct}{0.0}

\newcommand{\ResolvedEligiblePopulationNew}{4{,}277{,}878}
\newcommand{\RetractionCountFallbackNCitations}{205}
\newcommand{\RetractionCountFallbackNVenues}{42}
\newcommand{\RetractionMinRecentPublications}{10}
\newcommand{\RetractionModerateOrHigherCountBasedN}{205}
\newcommand{\RetractionModerateOrHigherN}{12{,}737}
\newcommand{\RetractionModerateOrHigherPct}{0.146}
\newcommand{\RetractionModerateOrHigherRateBasedN}{12{,}532}
\newcommand{\RetractionModerateOrHigherRateBasedPct}{98.4}
\newcommand{\RetractionNaiveEverRetractedPct}{13.16}
\newcommand{\RetractionNoteTierPctOfAnyHistory}{89.6}

\newcommand{\RetractionTinyDenominatorExampleNCitations}{1{,}815}
\newcommand{\RetractionTinyDenominatorExampleNPublications}{21{,}526}
\newcommand{\RetractionTinyDenominatorExampleRatePct}{0.0046}
\newcommand{\RetractionTinyDenominatorNVenues}{5}

\newcommand{\RwcloneExactStringMatchN}{370}
\newcommand{\RwcloneHijackedIssnMissingN}{14}
\newcommand{\RwcloneNTotal}{455}
\newcommand{\RwcloneOverlappingNormalizedMatchN}{65}
\newcommand{\SecondaryEligibleRatePct}{0.034}

\newcommand{\SisfIssnlBroadWithBothN}{5{,}928}

\newcommand{\SisfIssnlNLaterCacheOnly}{9}

\newcommand{\SisfIssnlNRefsCacheIssnDiffers}{46}
\newcommand{\SisfIssnlNRefsNullCacheOnly}{488}
\newcommand{\SisfIssnlNRestored}{8}

\newcommand{\SisfIssnlStrictWithBothN}{2{,}214}

\newcommand{\SisfSibNPromotedEntries}{2{,}199}
\newcommand{\SisfSibNPromotedFiles}{113}

\newcommand{\SisfSibNPromotedSubmissions}{93}

\newcommand{\StrictCount}{2{,}197}
\newcommand{\StrictIdentityUnverified}{784}
\newcommand{\StrictIdentityVerified}{33}
\newcommand{\StrictNSecondVoteSurvivors}{8}
\newcommand{\StrictNoIssn}{1{,}366}
\newcommand{\StrictNoIssnPct}{62.2}
\newcommand{\StrictUndecided}{14}

\newcommand{\StrictvoteGeResolvedCorrectedN}{13{,}396}

\newcommand{\StrictvoteVotesGeLegitResolvedCorrectedN}{156{,}767}

\newcommand{\SubbroadhistNFourplus}{110}
\newcommand{\SubbroadhistNOne}{3{,}940}
\newcommand{\SubmissionBroadPct}{2.82}
\newcommand{\SubmissionStrictPct}{1.15}

\newcommand{\UccdvNVerdictChanged}{36}

\newcommand{\UccdvNWithClonedVote}{236}
\newcommand{\UccdvStrictAfterN}{2{,}233}
\newcommand{\UccdvStrictBeforeN}{2{,}197}
\newcommand{\UnknownsensFivehundredthsCount}{3{,}332}
\newcommand{\UnknownsensFivehundredthsMultiple}{1.5}
\newcommand{\UnknownsensOnehalfCount}{13{,}545}
\newcommand{\UnknownsensOnehalfMultiple}{6.2}
\newcommand{\UnknownsensOnehundredthCount}{2{,}424}
\newcommand{\UnknownsensOnehundredthMultiple}{1.1}
\newcommand{\UnknownsensOnepercentCount}{24{,}894}
\newcommand{\UnknownsensOnepercentMultiple}{11.3}
\newcommand{\UnknownsensOnetenthCount}{4{,}467}
\newcommand{\UnknownsensOnetenthMultiple}{2.0}
\newcommand{\UnknownsensZeroCount}{2{,}197}
\newcommand{\UnknownsensZeroMultiple}{1.0}
\newcommand{\UnknownvenuesNDistinct}{198{,}248}
\newcommand{\UnknownvenuesNTotal}{2{,}269{,}673}
\newcommand{\UnknownvenuesTophundredPct}{56.0}
\newcommand{\UnknownvenuesToptenPct}{37.3}
\newcommand{\VenuetypeConferenceBroadPct}{0.007}

\newcommand{\VenuetypeJournalBroadPct}{0.125}